\documentclass[journal=langd5,manuscript=article]{achemso}
\setkeys{acs}{email=false}
\usepackage[english]{babel}
\usepackage[T1]{fontenc}
\usepackage{pdfpages}
\usepackage[fleqn]{amsmath}
\usepackage{graphicx}
\usepackage{hyperref}
\usepackage{subfigure}
\usepackage{cleveref}
\usepackage{verbatim}
\usepackage{enumerate}
\usepackage{blindtext}
\usepackage[utf8]{inputenc}
\usepackage{titlesec}

\title{}
\date{}

\begin{document}
\singlespacing
\vspace{-1.5in}
\begin{center}
	\begin{Large}
\textbf{Quantifying Confidence in DFT Predicted Surface Pourbaix Diagrams of Transition Metal Electrode-Electrolyte Interfaces}\\
	\end{Large}
\vspace{0.1in}
\begin{large}
\textbf{Olga Vinogradova$^{a}$, Dilip Krishnamurthy$^{b}$, Vikram Pande$^{b}$, Venkatasubramanian Viswanathan$^{a,b,*}$}\\
\end{large}
\end{center}
\vspace{0.1in}
$^a$ Department of Chemical Engineering, Carnegie Mellon University, Pittsburgh, Pennsylvania, 15213, USA\\
$^b$ Department of Mechanical Engineering, Carnegie Mellon University, Pittsburgh, Pennsylvania, 15213, USA\\
$^*$ Corresponding author, Email: venkvis@cmu.edu\\

\begin{abstract} 
Density Functional Theory (DFT) calculations have been widely used to predict the activity of catalysts based on the free energies of reaction intermediates. The incorporation of the state of the catalyst surface under the electrochemical operating conditions while constructing the free energy diagram is crucial, without which even trends in activity predictions could be imprecisely captured. Surface Pourbaix diagrams indicate the surface state as a function of the pH and the potential. In this work, we utilize error-estimation capabilities within the BEEF-vdW exchange correlation functional as an ensemble approach to propagate the uncertainty associated with the adsorption energetics in the construction of Pourbaix diagrams. Within this approach, surface-transition phase boundaries are no longer sharp and are therefore associated with a finite width. We determine the surface phase diagram for several transition metals under reaction conditions and electrode potentials relevant for the Oxygen Reduction Reaction (ORR).  We observe that our surface phase predictions for most predominant species are in good agreement with cyclic voltammetry experiments and prior DFT studies. We use the OH$^*$ intermediate for comparing adsorption characteristics on Pt(111), Pt(100), Pd(111), Ir(111), Rh(111), and Ru(0001) since it has been shown to have a higher prediction efficiency relative to O$^*$, and find the trend Ru>Rh>Ir>Pt>Pd for (111) metal facets, where Ru binds OH$^*$ the strongest. We robustly predict the likely surface phase as a function of reaction conditions by associating c-values to quantifying the confidence in predictions within the Pourbaix diagram. We define a confidence quantifying metric using which certain experimentally observed surface phases and peak assignments can be better rationalized. The probabilistic approach enables a more accurate determination of the surface structure, and can readily be incorporated in computational studies for better understanding the catalyst surface under operating conditions. 
\end{abstract}

\begin{tocentry}
\begin{center}
\includegraphics[width=0.9\linewidth,trim=1 1 20 20,clip]{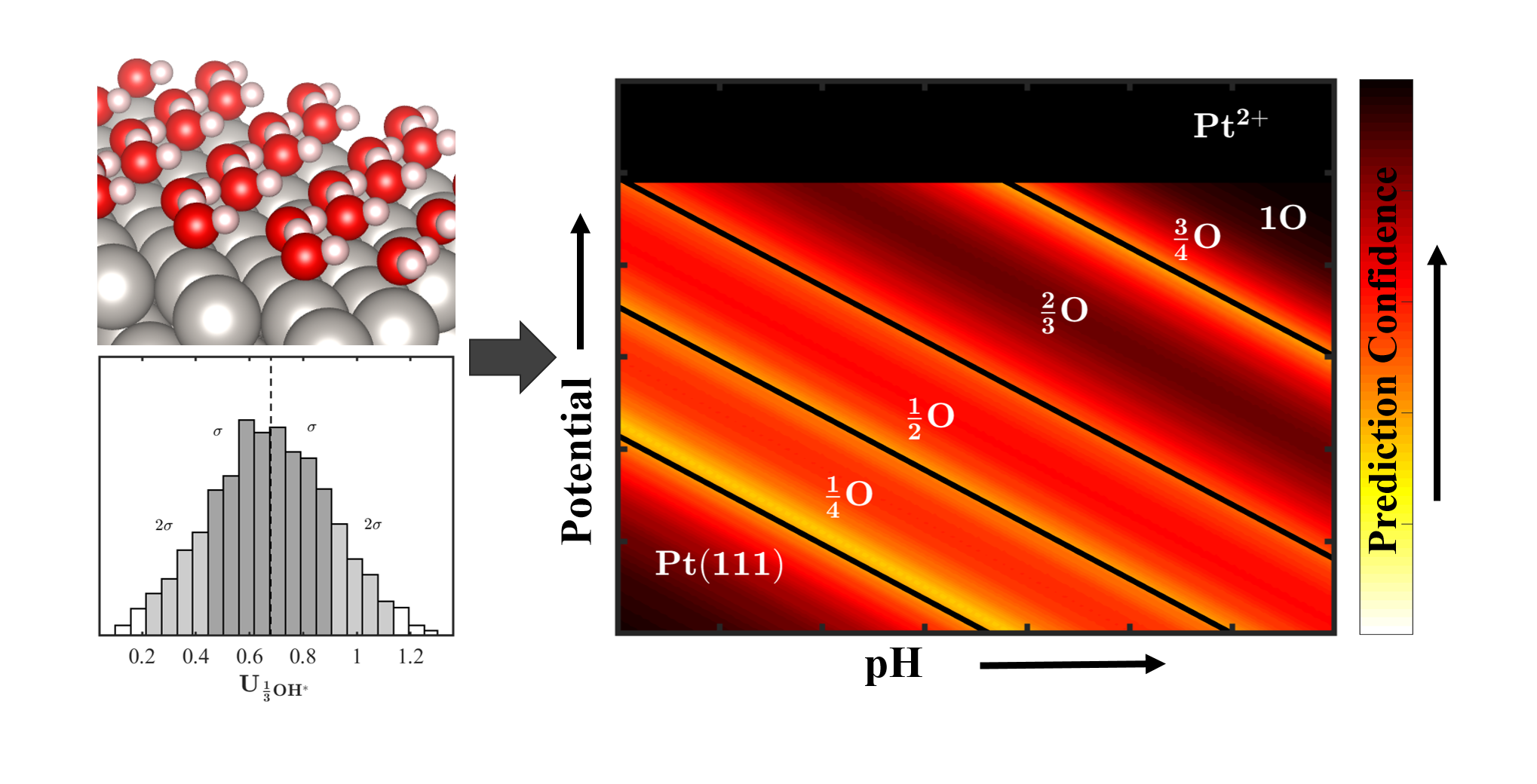}
\end{center}
\end{tocentry}

\section{Introduction}

Electrochemical processes are at the heart of many routes towards sustainable energy storage in the form of chemical bonds.\cite{lewis2006powering} The storage routes involve hydrogen and oxygen electrochemistry occurring at solid-liquid interfaces.\cite{montoya2017materials}  Determining the activity of electrode materials active for these processes requires a thorough understanding of the surface dynamics at the electrode-electrolyte interfaces.\cite{bockris2013surface}

Probing the electrode-electrolyte interface directly is an extremely challenging problem.\cite{bard1993electrode} There has been substantial development towards the coupling of X-ray spectroscopy directly with electrochemical cells.\cite{wakisaka2009identification,casalongue2013direct} This can be complemented by electrochemical impedance spectroscopy (EIS) to deconvolute the signals and assign them to surface adsorbed species.\cite{bondarenko2011pt} Theoretically, density functional theory calculations have been used to construct surface phase diagrams based on probing many different possible surface configurations.\cite{hansen2008surface}  However, a major challenge associated with this approach is that the choice of the exchange-correlation functional often plays a key role in determining the dominant surface phases.\cite{tian2009potentials,jinnouchi2008aqueous}

Recently, the use of Bayesian error estimation techniques has brought capabilities to estimate uncertainty associated with DFT simulations.\cite{wellendorff2012density}  In this work, we develop a systematic method to assign a confidence value (c-value) associated with the stable surface phases at a given electrode potential and pH.  Using this method, we probe the surface Pourbaix diagram of several transition metals: Pt, Pd, Ir, Rh, and Ru. We then map c-values onto the surface Pourbaix determined with the best-fit functional to gain insight into surface phase transitions and compare them with those determined by cyclic voltammetry experiments. Based on our method, we identify that on Pt(111) at an electrode potential of 0.65 V (vs. RHE) the surface could be covered with a combination of OH$^*$ or O$^*$ with relatively equal confidence.  Our method reproduces the well-known uncertainty associated with the DFT-predicted OH$^*$ to O$^*$ phase transition on Pt(111).\cite{rossmeisl2006calculated} On Pt(100), we note that the predicted stable phases of $\frac{1}{3}$OH$^*$ and  $\frac{1}{2}$OH$^*$ are consistent with the suggested structural transitions by Han et al.\cite{han2012first} Our results indicate that on Pd(111) at potentials starting at $\sim$0.75 V (vs. RHE), the surface is O$^*$ covered, consistent with the experimentally determined current peak corresponding to oxidation.\cite{hara2007preparation} On Ir(111), Rh(111), and Ru(0001) we find voltage regimes where our prediction confidence is limited relative to other surfaces explored in this work. The order of relative adsorption strengths of $\frac{1}{3}$OH$^*$ on (111) facet metals is Ru>Rh>Ir>Pt>Pd (in decreasing adsorption strength where Ru is most oxophilic). Except for the order between Pt and Pd, this ordering is comparable to the trend established in a previous computational study by Karlberg et al.\cite{karlberg2006adsorption} However, we observe that the adsorption strengths of $\frac{1}{3}$OH$^*$ on (111) facets of Pt and Pd are very close to each other (within 0.02 eV) demonstrating a need to consider prediction uncertainty.  We show that we can use c-value calculation to consider a wide array of functionals within DFT to assign an error to each adsorption energy. This is particularly helpful in distinguishing between surface phases close to each other in energy. Wider distributions of c-values correspond to greater uncertainties and vice versa. Comparing the widths of c-value energy distributions with cyclic voltammetry studies, we also consider a situation where the Rh(111) surface could reconstruct into metal-oxides to explain certain features in cyclic voltammograms. Given the importance of the stable state of the surface in determining electrochemical reaction mechanisms, we believe this method will be extremely crucial in constructing free energy diagrams on the appropriate surface and will play an important role in determining the subsequent limiting potential.\cite{deshpande2016quantifying,krishnamurthy2018maximal}

\section{Methods}
\subsection{Construction of Surface Pourbaix Diagrams}
\label{sec:surface_pourbaix}
 The surface Pourbaix diagram represents the stable state of the catalyst surface in electrochemical systems as a function of pH and the electrode potential $U$. Consider an arbitrary metal surface $S$ with adsorption site $*$ in its pristine state. A generalized representation of the adsorption of oxygenated intermediates, denoted as $\mathrm{O_mH_n^*}$, can be written as 

\begin{equation}
    \mathrm{S\mbox{-}O_mH_n^* + (2m-n)(e^- + H^+) \rightleftharpoons S^* +  m H_2O}
\label{eqn:equilibrium}
\end{equation}

where $m$ and $n$ are the number of oxygen and hydrogen atoms in the adsorbate respectively. Note that this representation corresponds to a concerted proton-coupled electron transfer (PCET) reaction. The associated free energy change can be computed as:
\begin{equation}
    \mathrm{\Delta G(U,pH) = G_{S^*} + mG_{H_2O} -  G_{S\mbox{-}O_mH_n^*} - (2m-n)(G_{e^-} + G_{H^+}) }
\label{eqn:G1}
\end{equation}


Rewriting Eqn. \ref{eqn:G1}, the free energy of electrons $G_{e^-}$ is defined as $-eU$ where $e$ is the electron charge. The free energy of protons $G_{H^+}$ is determined from the computational hydrogen electrode using the equilibrium relation $H^+ + e^- \rightleftharpoons H_2$. This gives at non-standard conditions $G_{H^+}+G_{e^-} = \frac{1}{2}G_{H_2} - eU_{SHE} + k_BT(\ln[a_{H^+}])$ where $U_{SHE}$ is the potential relative to the standard hydrogen electrode (SHE). We then write the free energy change for adsorption of $\mathrm{O_mH_n^*}$ intermediates as 
\begin{equation}
    \mathrm{\Delta G(U,pH) = G_{S^*} + mG_{H_2O} - G_{S\mbox{-}O_mH_n^*} - (2m-n)(\frac{1}{2}G_{H_2}  - U_{SHE} - 2.303k_BTpH) } 
\label{eqn:G2}
\end{equation}
Therefore, we can derive a relation between potential and pH for a wide variety of adsorbates on a surface $S$ in reference to standard conditions when $\Delta G(U,pH)=0$. 

Similarly, generalized pH independent electrochemical reactions such as the dissolution of metal surfaces can be represented as: 
\begin{equation}
    \mathrm{S^{i+} \rightleftharpoons S^{j+}+(j-i)e^- }
\label{eqn:equilibrium3}
\end{equation}

where $i$ and $j$ represent the charge on the ion, which can be zero for metal surfaces and uncharged species. The free energy would therefore be pH-independent and appear as a horizontal lines on the surface Pourbaix and the standard reduction potentials are taken from the CRC handbook.\cite{lide1994crc} We note that unconcerted electrochemical reactions can be represented as linear combinations of the two generalized forms presented above.

Since the most stable state at a given condition is that which minimizes free energy, we determine the surface Pourbaix diagram by choosing the appropriate state at each set of conditions of U and pH. The phase transitions between adsorbates appear as lines with a slope of -59mV/pH at standard conditions. Free energies are determined from calculated energies of adsorption $E_{S^*}$ corrected for entropy, S, and zero point energies, ZPE, which we assume do not change significantly with temperature, by $\mathrm{\Delta G=\Delta E_{ref,water}-T\Delta S+\Delta ZPE}$. Values for entropy and zero point energy correction are given in the Supporting Information section 5. In this work, we assume constant zero point energies across the ensemble of functionals for simplicity, since it is expected to have negligible variation (of order $kT$, where $k$ is the Boltzmann constant) relative to the order of magnitude ($10-20$ $kT$) of the surface energetics.

\subsection{Bayesian Error Estimation Framework}
 With the recent development of the Bayesian Ensemble Error Functional with van der Waals correlations (BEEF-vdW),~\cite{wellendorff2012density} there exists a systemat approach to estimate the uncertainty in energetics of a DFT calculation. BEEF-vdW is a semi-empirical exchange correlation (XC) functional including non-local  contributions, developed by using training data sets for molecular formation energies, molecular reaction energies, molecular reaction barriers, non-covalent interactions, solid state properties, and chemisorption on solid surfaces. 
	
The exchange correlation energy in BEEF-vdW is expressed as a sum of the GGA exchange energy expanded using Legendre polynomials, the LDA and PBE\cite{perdew1996generalized} correlation energies and the non local correlation energy from vdW-DF2.\cite{lee2010higher}
\begin{equation}\label{eq:1}
\mathrm{E_{xc}=\sum_{m} a_m E_m^{GGA-x} + \alpha_c E^{LDA-c}+(1-\alpha_c)E^{PBE-c}+E^{nl-c}}
\end{equation}
The parameters $a_m$ and $\alpha_c$ are optimized with respect to the above mentioned data sets. The error estimation functionality is enabled by deriving an ensemble of energies from an ensemble of exchange correlation functionals non-self-consistently following a self-consistent DFT calculation. The ensemble of exchange correlation functionals is generated using a probability distribution function for the parameters $a_m$ and $\alpha_c$ such that the standard deviation of the ensemble of energies reproduces the standard deviation for the training properties calculated using BEEF-vdW self-consistently.

\subsection{Prediction Confidence of Surface Pourbaix Diagrams}
\label{sec:cvalue_pourbaix}
Pourbaix diagrams (figure \ref{fig:Pt111_Pourbaix}) typically depict only the most stable state of the surface (minimum Gibbs free energy) over a range of operating potentials (U) and pH values with sharp phase boundaries. We utilize error estimation capabilities within the BEEF-vdW exchange correlation functional to determine the ensemble of free energies for all considered states as a function of U and pH. Therefore, for every functional there exists a unique set of stable surface states in the ranges of pH and U of interest, determined by the corresponding minimum free energy surface state. This results in an ensemble of Pourbaix diagrams, allowing us to apply statistical tools to obtain a measure of the confidence in a predicted surface state by quantifying the agreement between functionals. We are interested in determining the level of agreement between functionals at the GGA-level as to the most energetically favorable surface state as a function of potential and pH. We use the confidence-value\cite{sumaria2018quantifying,houchins2017quantifying} (c-value), which in this context is defined as the fraction of the ensemble that is in agreement with the hypothesis of the optimal BEEF-vdW (best-fit) functional, and
is given by


\begin{equation}
c(U,pH) =  \frac{1}{N_{ens}} \sum_{n=1}^{N_{ens}} \prod_{s_i\neq s_{opt}} \Theta(\Delta G_{s_i}^n(U,pH)-\Delta G_{s_{opt}}^n (U,pH))
\end{equation}
 where, $N_{ens}$ is the number of functionals in the ensemble, which is $2000$ in this work. $\Theta(x)$ denotes the Heaviside step function. At any given $U$ and $pH$, $s_i \in S$, the set of all considered surface states, and $s_{opt}$ is the thermodynamically stable surface state predicted by the BEEF-vdW optimal functional. $\Delta G^n_{s_i}$ refers to the free energy of the $i^{\mathrm{th}}$ surface state given by the $n^{\mathrm{th}}$ member of the ensemble of functionals. This approach allows assigning confidence values to calculated surface Pourbaix diagrams.
 
 In regimes of electrode potential and pH where the confidence value is lower than 1, it is important to determine whether the stable surface phase predicted by the majority of functionals agree with the hypothesis of the best-fit functional. We extend the concept of c-value to determine a measure of confidence in any considered surface phase being the lowest in free energy, looking beyond just agreement with respect to the most stable phase in accordance with the optimal functional hypothesis. In order to define this measure, we use $sp(U,pH)$ as the function that maps electrochemical conditions to the corresponding most stable surface phase ("sp" for "stable phase") from the set of possible/considered phases denoted by \{0,1,2,...,i,...,n\}, where $i$ denotes the $i^{\mathrm{th}}$ surface phase and $i=0$ indicates the clean surface. For example, for the $i^{\mathrm{th}}$ considered surface phase, we define below the respective stability confidence measure $c_{sp=1}(U,pH)$ quantitatively as the fraction of functionals that predict that the phase labelled as $i=1$ is the lowest in free energy:
\begin{equation}
\label{spredi}
c_{sp=i}(U,pH) = \frac{1}{N_{ens}} \sum_{n=1}^{N_{ens}}\delta (sp^n(U,pH)-i).
\end{equation}
where $\delta(x)$ denotes the Dirac delta function and the superscript $n$ denotes the prediction from the $n^{\mathrm{th}}$ functional.



\section{Results and Discussion}
We use the example case of the surface phase diagrams relevant for oxygen reduction reaction (ORR) on Pt-group transition metals: Pt, Pd, Ir, Rh, and Ru, to demonstrate an approach to quantify the ability of DFT in determining the state surface state.  We present our findings and compare them to experimental linear sweep voltammetry measurements, and discuss the degree of confidence in our DFT predictions in relation to the assignment of voltammetric peaks.  

\subsection{Pt(111)}

\begin{figure}
	\centering
	\includegraphics[width=0.7\linewidth]{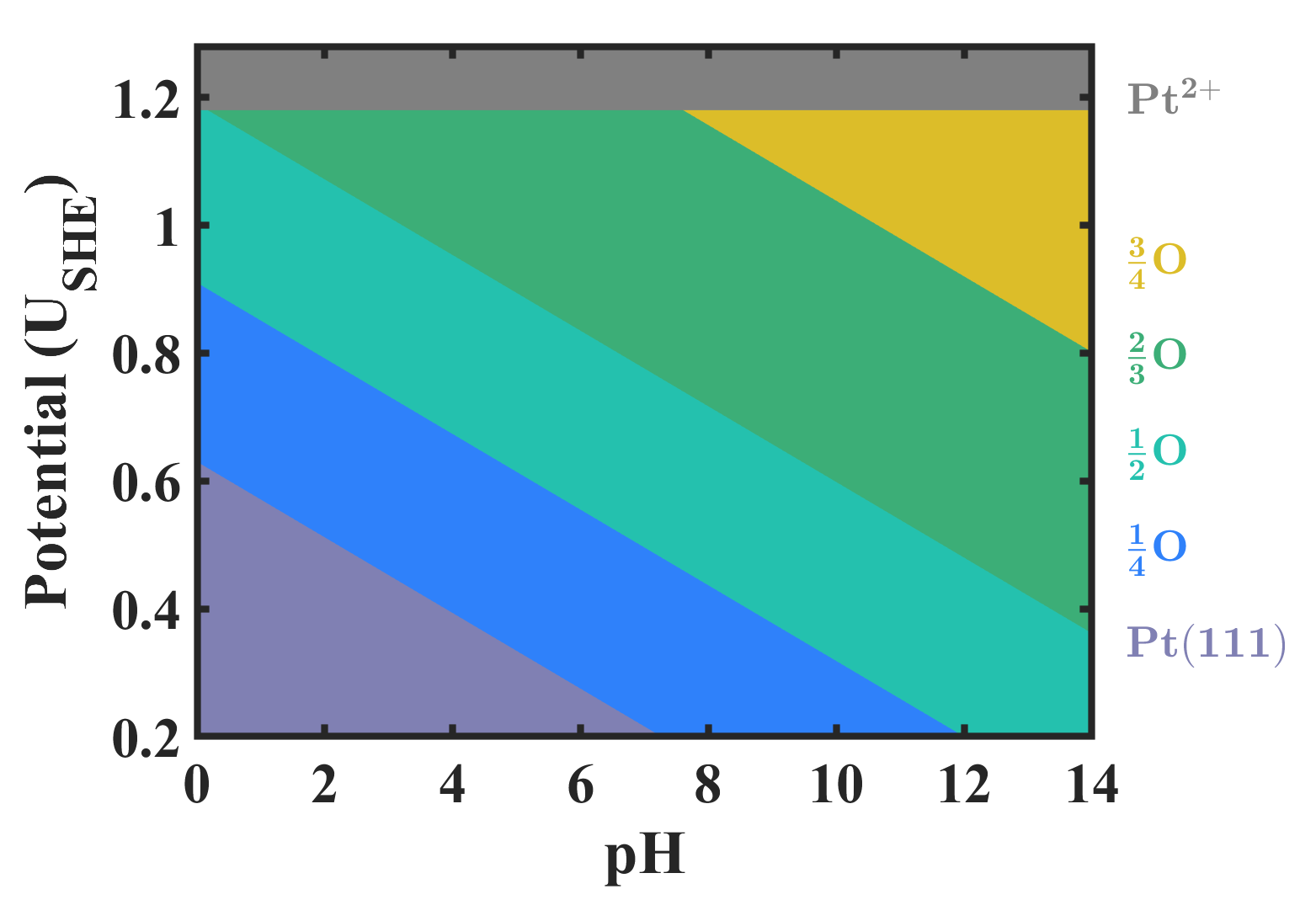}
	\caption{For Pt(111) the surface Pourbaix diagram shows regions of stability for surface coverages of OH$^*$ and O$^*$ at conditions of pH and potential relative to the standard hydrogen electrode (SHE). Dissolution of Pt is shown at 1.18 V and is independent of pH.  }
	\label{fig:Pt111_Pourbaix}
\end{figure}

\begin{figure}
	\centering
	\includegraphics[width=0.7\linewidth,trim=1 1 20 20,clip]{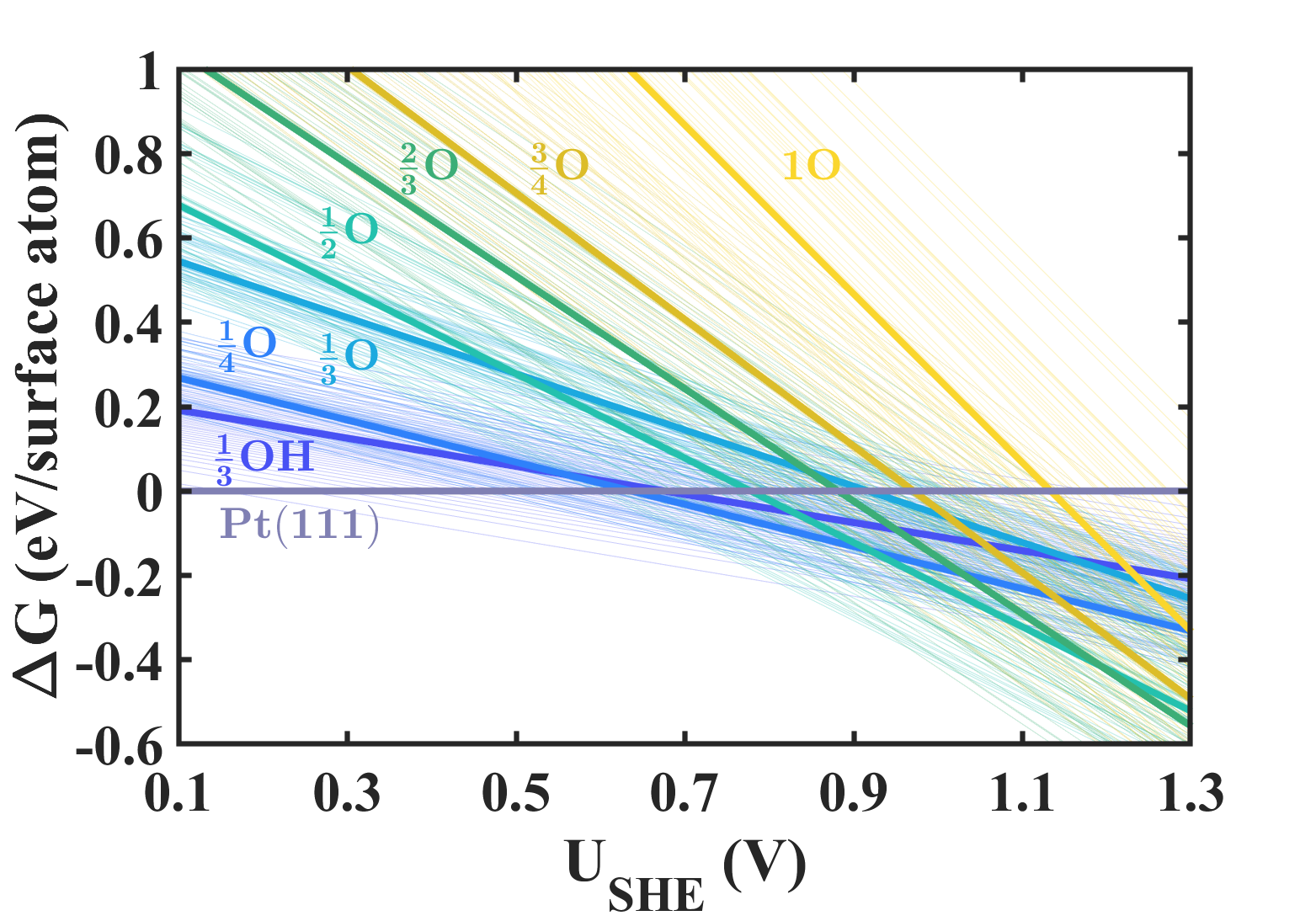}
	\caption{Comparing free energies of all coverage fraction surfaces sampled for Pt(111) at pH=0 we observe from the best-fit solutions (thick lines) that 1/4 O$^*$ is <0.05 eV more stable than 1/3 OH$^*$ at 0.63 V. The thin lines represent solutions for free energy for a sample set of 50 BEEF-vdW ensemble functionals depicting a spread of energies for each surface state determined using the best-fit functional.  }
	\label{fig:Pt111_GvsU}
\end{figure}

\begin{figure}
	\centering
	\includegraphics[width=1.0\linewidth,trim=1 1 20 20,clip]{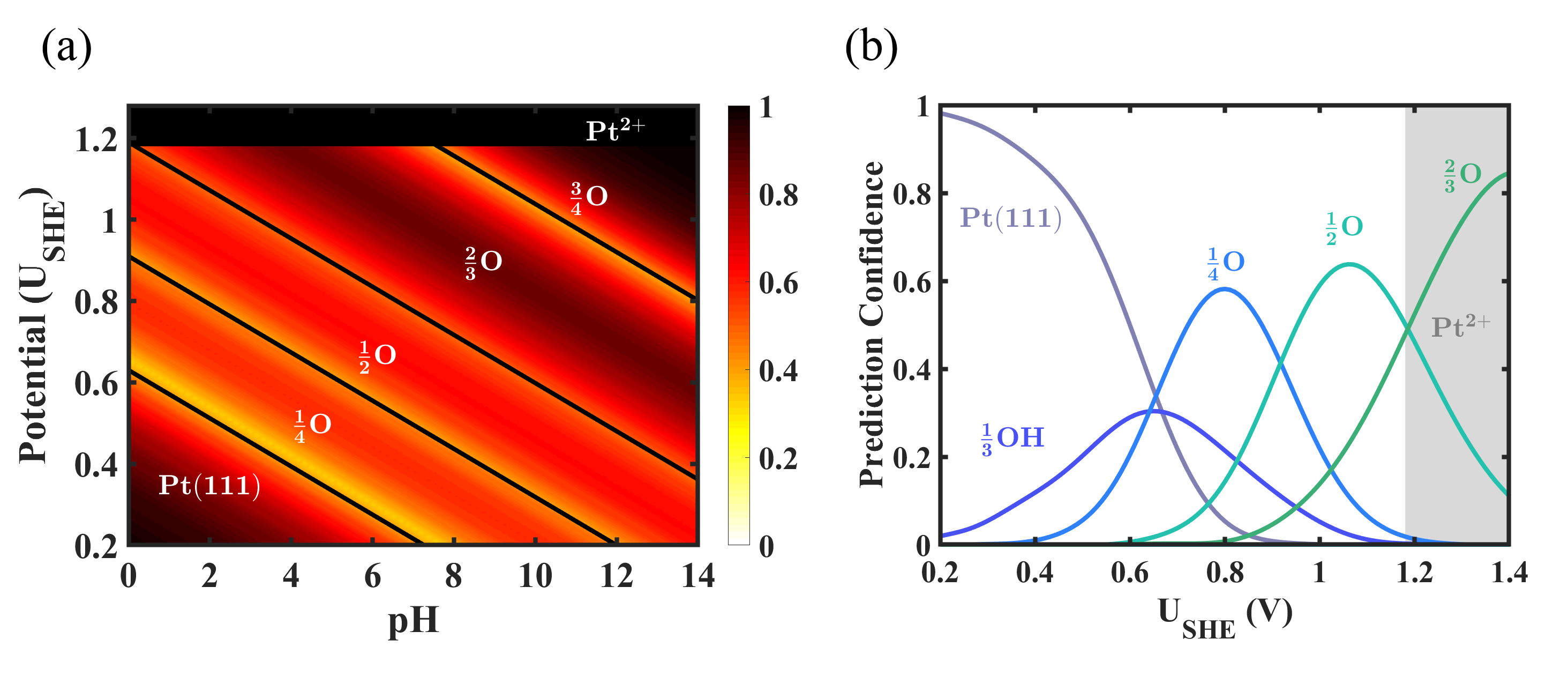}
	\caption{Figure (a) shows the prediction confidence of each stable surface of a Pourbaix diagram for Pt(111). Greatest prediction uncertainty occurs at phase transition boundaries. In (b) we plot the prediction confidence in relation to potential at pH of 0. Surface states with the largest c-values correspond to the solution determined with the best-fit BEEF-vdW XC-functional in figure \ref{fig:Pt111_Pourbaix}. We also note that this representation of state surfaces may be extended to higher pH values (refer to Supporting Information section 8.}
	\label{fig:Pt111_2}
\end{figure}

The Pourbaix diagram for Pt(111), previously investigated\cite{rossmeisl2006calculated,hansen2008surface}, is calculated using the BEEF-vdW exchange correlation functional and is shown in  figure \ref{fig:Pt111_Pourbaix}. We consider the following oxygenated adsorbate surface states of Pt(111): 1/3 OH$^*$, 1/4 O$^*$, 1/3 O$^*$, 1/2 O$^*$, 2/3 O$^*$, 3/4 O$^*$, and 1 O$^*$. While modeling OH$^*$ with water stabilization effects, we ignore potential dependent water orientation effects\cite{garcia2009potential}, and we neglect water stabilization for O$^*$ states\cite{ogasawara2002structure} as both of these effects have been shown to have a negligible influence on the energetics. Locally mixed surface phases are expected to be energetically unfavorable and are not considered since surface O$^*$ repels water unlike adsorbed OH$^*$, where a hexagonal water stabilizing layer occurs due to favorable hydrogen bonding.\cite{ogasawara2002structure,tian2009potentials} In addition, mixed phases are complex to simulate\cite{rossmeisl2006calculated} since it involves the incorporation of long range disorder, which is undoubtedly present.\cite{viswanathan2012simulating} We also neglect configurational entropy of adsorbed oxygen intermediates, since the effect is expected to be small.\cite{rossmeisl2009steady}

The state of the surface is governed by the following water discharge reactions where * refers to an adsorption site on a metal catalyst surface:
\begin{equation*}
\mathrm{H_2O + *\rightleftharpoons OH^{*} + e^{-} + H^{+}}
\end{equation*}
\begin{equation*}
\mathrm{H_2O + *\rightleftharpoons O^{*} + 2e^{-} + 2H^{+}}    
\end{equation*}

 We compute the free energy changes of these reactions, which are functions of the pH and the electrode potential, based on Eqn. \ref{eqn:G2} (refer section \ref{sec:surface_pourbaix}).
\begin{equation}
\mathrm{\Delta G = (G_{OH^*}+\frac{1}{2}G_{H_2}-G_{H_2O}) + {k_BT} ln[a_{H^+}]-eU}
\label{f1}
\end{equation}
\vspace{-1cm}
\begin{equation}
\mathrm{\Delta G = \frac{1}{2}(G_{O^*}+G_{H_2}-G_{H_2O}) + {k_BT} ln[a_{H^+}]-eU}
\label{f2}
\end{equation}

The surface Pourbaix diagram for Pt(111) shown in figure \ref{fig:Pt111_Pourbaix} is constructed using Eqns. \eqref{f1} and \eqref{f2} . The free-energy as a function of the electrode potential (figure \ref{fig:Pt111_GvsU}) determines the surface phase boundaries from the best-fit BEEF-vdW XC functional and therefore identifies the most stable state of the surface at given reaction conditions of potential and pH.  At proton activity of a$_{H^+}$=1, we predict that below $\approx$0.63 V (vs. RHE), clean Pt(111) surface is the most stable. Water oxidation occurs at higher potentials starting with 1/4 monolayer coverage of O$^*$ and increasing monotonically to coverage 1/2 O$^*$ at $\approx$0.91 V (vs. RHE). Near 1.18 V, we indicate Pt dissociation\cite{lide1994crc} through $Pt \rightleftharpoons Pt^{2+} + 2 e^{-}$, which is pH independent and appears as a horizontal line on the surface Pourbaix diagram. Although there exist some differences in the positions of the phase boundaries, the observed set of stable surface states are consistent with those predicted by Hansen et al\cite{hansen2008surface} using the RPBE XC-functional with the exception of the 1/3 OH$^*$ surface state which Hansen et al. predicts at 0.78 V (vs. RHE). However, we observe that in the range $0.63$--$0.83$ V (vs. RHE) the 1/4 O$^*$ surface is marginally (by $<0.05$ eV) more stable than the 1/3 OH$^*$ surface state, an observation also noted by Hansen et al. in the region $0.78$--$0.84$ V. Close energetics can be resolved more robustly through the quantification of confidence in the relative stability of surface phases by the use of error estimation capabilities based on agreement between functionals at the GGA level. Using the BEEF-vdW best-fit functional, we compute the equilibrium potential of the 1/3 OH$^*$ surface state to be 0.68 V (vs. RHE). Since the ensemble of functionals within the BEEF-vdW family of functionals give rise to an ensemble of energies, we quantify the uncertainty by the standard deviation of the distribution, computed to be $\sigma_{OH}$=0.23, which bounds within the 68$\%$ confidence interval the calculation of 0.81 V by Rossmeisel et al\cite{rossmeisl2006calculated} using the RPBE XC-functional. Other examples of calculations bound within one standard deviation of the ensemble approach include the OH$^*$ adsorption energy computed by Taylor et al\cite{taylor2007first} to be 0.59 V for a lower coverage of 1/9 OH$^*$ using the PW91 XC-functional, and the adsorption potential for 1/3 OH$^*$ to be 0.71 V (PBE XC-functional) and 0.69 V (RPBE XC-functional) by Jinnouchi et al.\cite{jinnouchi2008aqueous} Similarly, calculations using RPBE XC-functional by Anderson et al. measured adsorption of a variety of combinations of OH$^*$ and H$_2$O fractions, observing that adsorbed OH$^*$ begins to form at $\sim$0.6 V.\cite{tian2009potentials}

The ensemble of functionals within the BEEF-vdW XC functional results in an ensemble of free energies for each state (figure \ref{fig:Pt111_GvsU}) which gives rise to an ensemble of Pourbaix diagrams.\cite{sumaria2018quantifying} Using c-value to quantify confidence in the predicted surface surface phases based on the fraction of the ensemble that is in agreement with the hypothesis of the best-fit (or optimal BEEF-vdW) functional, we construct a modified surface Pourbaix diagram following the methodology described in section \ref{sec:cvalue_pourbaix}.  The confidence-derived surface Pourbaix diagram of Pt(111) with the associated c-values is shown in figure \ref{fig:Pt111_2}a. We note that regions spanning  surface phase transitions are those with the lowest c-values. To better understand the relative confidence values, we show the c-values of each surface phase as a function of potential at fixed pH value of 0 in figure \ref{fig:Pt111_2}b. In this figure, we plot  c$_{sp=i}(pH,U)$ (refer to eqn. \ref{spredi} of section \ref{sec:cvalue_pourbaix}) for all $i$ corresponding to the various considered surface phases to capture the degree of agreement between functionals as to the lowest free energy phase.

The phase transitions in figure \ref{fig:Pt111_2}b predicted by the $max(c_{sp=i}(U))$ curve at fixed pH values lie within 0.02 eV of those predicted by the optimal BEEF-vdW XC-functional. We observe that the distribution of confidence values for OH$^*$ is wider than that for the O$^*$ phases, and attribute this to a larger standard deviation in the distribution of energies for water-stabilized OH$^*$ ($\sigma_{OH}$=0.23 eV) compared to O$^*$ ($\sigma_{O}\approx$ 0.1 eV) in reference to a water layer. The distributions of the free energies of adsorption have similar widths, however, in calculating the equilibrium potential, O$^*$ is associated with the concerted transfer of two protons and electrons while the formation of OH$^*$ involves only one proton-coupled electron transfer (Eqns. \ref{f1} and \ref{f2}). The additional factor of 2 provides rationale for the tightening of the equilibrium potential distribution for O$^*$ relative to OH$^*$. We discuss the relative confidence values of the various phases at pH=0, however at different pH values the effect can be captured by a constant chemical potential shift of the proton activity (see Supporting Information section 8). At U$_{RHE}$ < 0.55 V, we predict the clean Pt(111) surface with the highest c-value ($>0.5$). Around $\approx$0.63 V (RHE), we find that the confidence values of the clean Pt(111), 1/3 OH$^*$ and 1/4 OH$^*$ are nearly equal ($\approx 0.33$) indicating that within DFT at the GGA-level these surface phases are equally likely to be the lowest in free energy. The uncertainty in the OH$^*$ to O$^*$ phase transition in the range from $\sim$0.6 V to $\sim$0.8 V (RHE) is consistent with that reported earlier.\cite{casalongue2013direct} We note that in linear sweep voltammetry experiments Wang et al\cite{wang2004kinetic}, Koper\cite{koper2011blank}, and Garcia-Araez et al\cite{garcia2008determination,garcia2006thermodynamic} attribute the reversible butterfly region between 0.6 V and 0.85 V  to OH$^*$, which is also supported by Wakisaka et al\cite{wakisaka2009identification} in ex situ XPS experiments. In line with these findings, although at $U\approx$0.63 V (RHE) predictions from our analysis (figure \ref{fig:Pt111_2}b) have very close energetics, we note that the highest prediction confidence for OH$^*$ adsorption occurs at this potential, a key observation that is missed from the surface Pourbaix diagram in figure \ref{fig:Pt111_Pourbaix}. This emphasizes the usefulness of the defined quantity $c_{sp=i}(U)$ in more robustly assigning electrode potential regimes for the stability of a given surface state, when close energetics can be resolved between surface phases through experimental validation or higher-order DFT. At U$_{RHE}$ > 0.63 V (RHE), O$^*$ surface adsorption coverage gradually increases to higher oxygen coverage and higher c-values for 1/4 O$^*$ and 1/2 O$^*$ until  dissolution at a potential of 1.18 V. The 2/3 O $^*$ surface phase is not stable at potentials below the dissolution potential. Wakisaka et al.\cite{wakisaka2009identification} have reported that OH$^*$ formation begins at E > 0.6 V continuing to increase until 0.8 V which is captured within uncertainty bounds of our predictions. Our findings of a monotonic increase in the oxygen coverage compare well with blank voltammetry\cite{bondarenko2011pt} of Pt(111) in 0.1 M HClO$_4$, avoiding the specific adsorption of sulfate that occurs in sulfuric acid electrolyte. We believe that the correlation between our predictions and state-of-the-art experiments strengthens the effectiveness of confidence values and Bayesian error estimation tools in quantifying uncertainty of predicted stability of surface phases. 

\subsection{Pt(100)}

\begin{figure}
	\centering
	\includegraphics[width=15cm,,trim=1 1 20 20,clip]{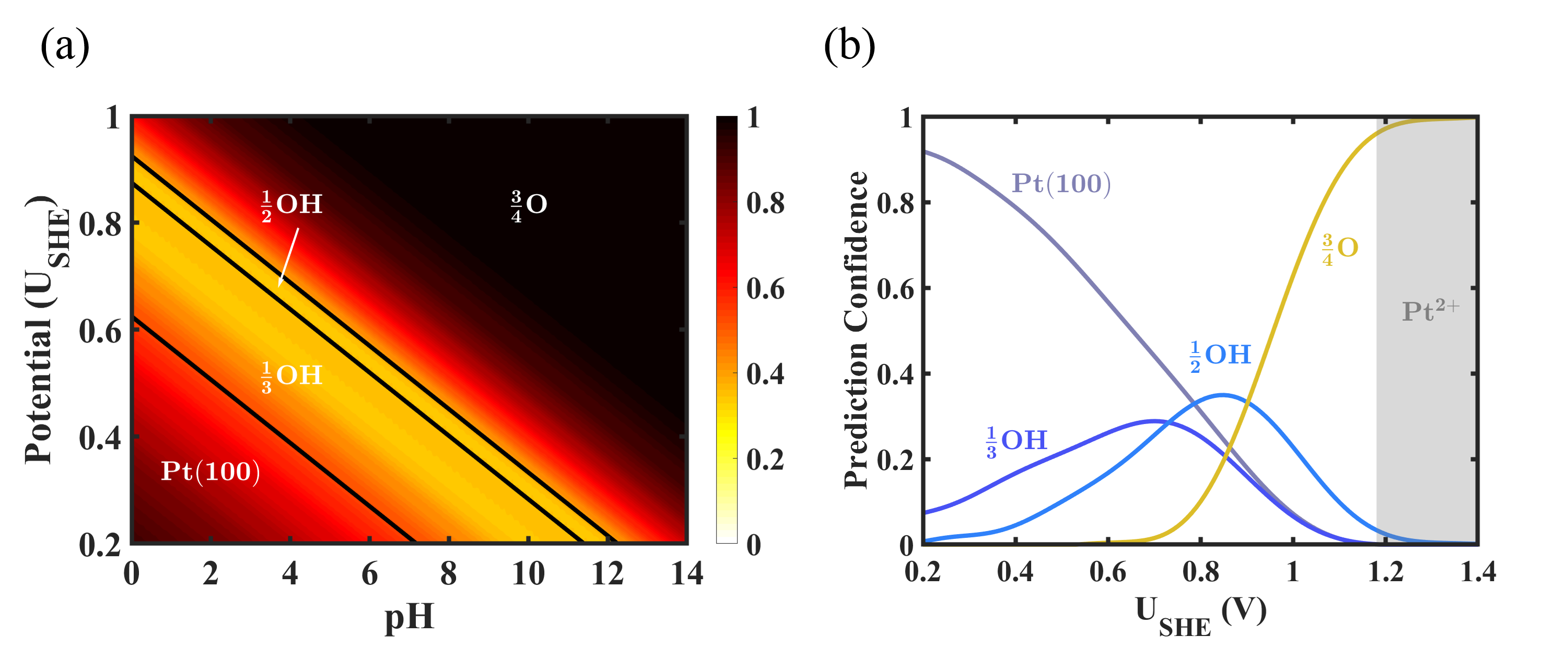}
	\caption{Figure (a) shows the surface Pourbaix diagram for Pt(100) where the onset of 1/3 OH$^*$ is at 0.64 V (vs. RHE). A free-energy vs potential figure is presented in the Supporting Information section 9.1. Figure (b) depicts the c-value for each surface state at pH=0. We observe a gradual transition from 1/3 OH$^*$ to 1/2 OH$^*$ from $\sim$0.6 V to $\sim$0.9 V (RHE). At U$_{RHE}$ > 0.9 V there is greatest likelihood of observing 3/4 O$^*$. The region on the x-axis shaded in gray represents the dissolution potential occurring at 1.18 V for Pt.}
	\label{fig_Pt100}
\end{figure}

We apply a similar methodology to Pt(100), which has been well studied both experimentally\cite{stamenkovic2001structure,climent2006thermodynamic,gomez2004effect} and computationally\cite{duan2013comparison,han2012first} since commercial state-of-the-art Pt catalysts are comprised typically of low-index (111) and (100) facets\cite{song2005pt,wang2008general}. We consider the following surface states of Pt(100): 1/3 OH$^*$, 1/2 OH$^*$, 1/4 O$^*$, 1/2 O$^*$, 3/4 O$^*$ and 1 O$^*$. The surface phase boundaries in figure \ref{fig_Pt100}a are determined using the best-fit BEEF-vdW functional. On the surface Pourbaix diagram we overlay the c-values as a function of pH and potential, which provide a measure of agreement between GGA-level functionals in the BEEF-vdW family of functional. We find that at U$_{RHE}$ < 0.64 V, the clean Pt(100) surface is thermodynamically most favorable. At higher potentials we predict 1/3 OH$^*$ until 0.87 V (RHE) at which there is a narrow potential window where the 1/2 OH$^*$ phase is the lowest in free energy. At U$_{RHE} \approx >$ 0.9 V  the OH$^*$ phase transitions to the 3/4 O$^*$ surface. We observe that the prediction confidence (c-value) of the OH$^*$ surface states is relatively low ($\approx$<0.4). The narrow region for stable 1/2 OH$^*$ suggests that precise determination of the phase boundaries for this state is computationally challenging with the best-fit BEEF-vdW functional. We show the degree of agreement within the ensemble of functionals as to the most stable surface phase for each individual surfaces state in figure \ref{fig_Pt100}b. We see that in the range 0.6 V<U$_{RHE}$<0.7 V 1/3 OH$^*$ has a marginally greater c-value than 1/2 OH$^*$. However, at $\approx$0.8 V (RHE) the 1/2 OH$^*$ surface state has a c-value of $<0.25$ suggesting that the 1/3 OH$^*$ phase transitions to the more thermodynamically stable 1/2 OH$^*$ with increasing electrode potential. Previously it was suggested through ab-initio DFT calculations and Monte Carlo simulations\cite{han2012first} that 1/3 OH$^*$ transforms into 1/2 OH$^*$ which is supported by our analysis; we find that 1/2 OH$^*$ has the greatest confidence between 0.75 V and 0.9 V (vs. RHE). Of all oxygen covered surfaces considered, at U$_{RHE}$ > 0.9 V we predict the 3/4 O$^*$ adsorption state to be the most favorable until the dissolution potential. Based on cyclic voltammetry in hydrochloric acid performed by several groups\cite{climent2006thermodynamic,gomez2004effect} a wide reversible peak centered at 0.36 V (vs. RHE) has been reported, and attributed to a strong overlap between hydrogen and OH$^*$ adsorption regions which makes precise determination of adsorbed states difficult. Gomez et al\cite{gomez2004effect} used a deconvolution analysis to decouple OH coverage from H coverage, and in a subsequent work a maximum coverage value of 0.37 OH species per Pt surface atom was reported\cite{climent2006thermodynamic}, which corresponds to OH$^*$ adsorption assigned to a shoulder of the reversible peak appearing between 0.5 V and 0.7 V on the cyclic voltammograms. We highlight that the prediction from our analysis for the clean Pt(100) surface transition to OH$^*$ (1/3 OH$^*$ at 0.65 V (vs. RHE)) falls within this region although the relatively low ($<0.4$) confidence values indicate the uncertainty associated with this transition. Similar to the case of Pt(111), this again emphasizes the efficacy of the defined quantity $c_{sp=i}(U)$ in assigning potential regimes for the stability of a given surface state, when close energetics can be resolved between surface phases through insights from both theory and experiments. We do not consider H$^*$ adsorption since it is beyond the scope of the current work owing to the necessary coverage dependent analysis.\cite{karlberg2007cyclic}

\subsection{Pd(111)}

\begin{figure}
	\centering
	\includegraphics[width=15cm,trim=1 1 20 20,clip]{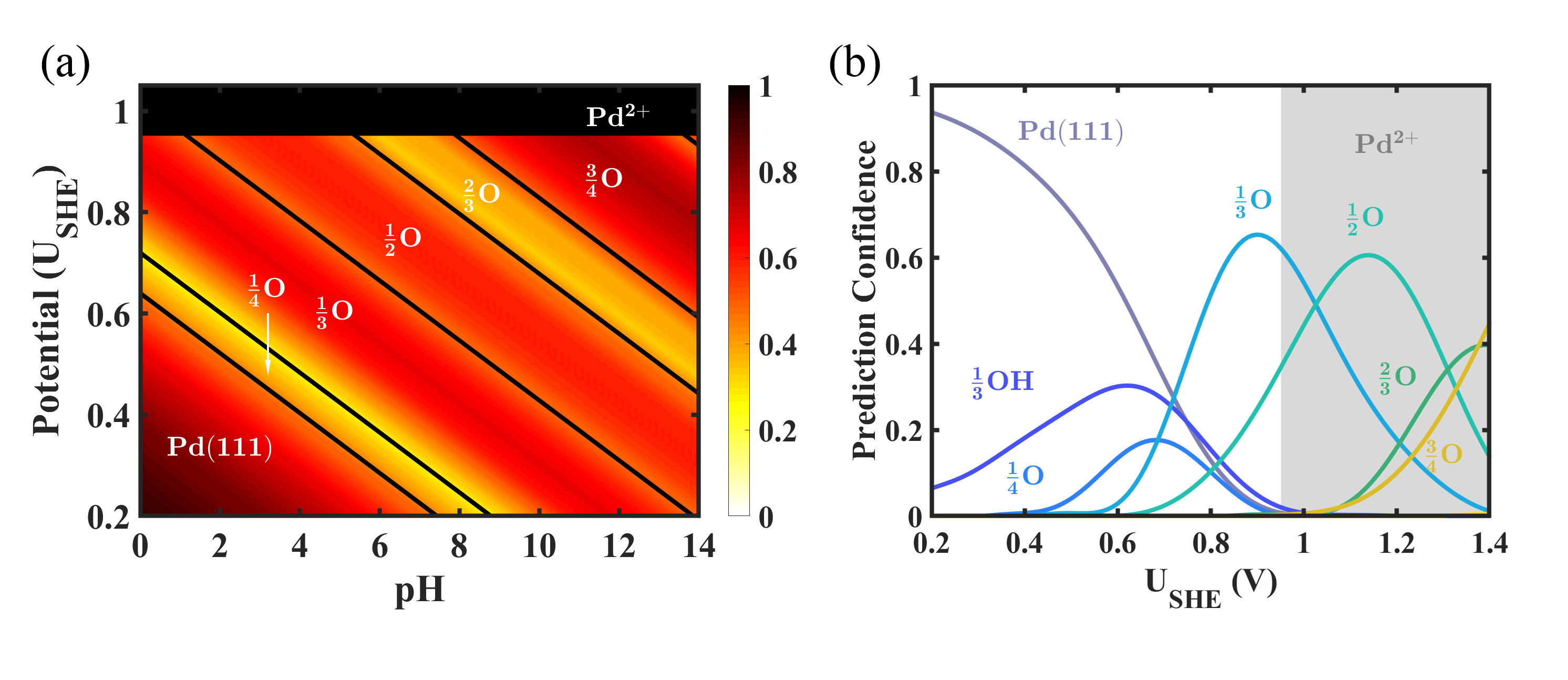}
	\caption{Figure (a) shows the surface Pourbaix diagram for Pd(111), where at pH=0 oxidation is predicted to begin at 0.64 V (vs. RHE). Figure (b) shows the c-value for each surface state at pH=0, where transition to oxygen covered surfaces begins at U > 0.7 V. The shaded gray region represents the onset of dissolution of crystalline Pd at 0.95 V.}
	\label{fig_Pd111}
\end{figure}

Palladium has been identified to be a promising alternative to Pt for the ORR\cite{shao2006pd,fernandez2005pd, fernandez2005thermodynamic,savadogo2004new}. We construct the Pourbaix diagram for Pd(111) by considering the following states of the surface\cite{mitsui2002water} similar to those considered for Pt(111): clean surface, 1/3 OH$^*$, 1/4 O$^*$, 1/3 O$^*$, 1/2 O$^*$, 2/3 O$^*$, 3/4 O$^*$, and 1 O$^*$. We predict from the constructed surface Pourbaix diagram, shown in figure \ref{fig_Pd111}a, that the surface is clear of adsorbates until a potential of 0.64 V (vs. RHE) which marks the onset of 1/4 O$^*$ coverage surface state. At higher electrode potentials, O$^*$ adsorption increases monotonically to 1/3 O$^*$ followed by 1/2 O$^*$, 2/3 O$^*$, and finally by 3/4 O$^*$. $Pd \rightleftharpoons Pd^{2+} + 2 e^{-}$ dissolution is labeled at 0.95 V which is independent of pH. Relatively low ($<0.4$) prediction confidence regions in figure \ref{fig_Pd111}a show that transitions between 1/4 O$^*$ and 1/3 O$^*$ in particular have a high degree of uncertainty along with the 2/3 O$^*$ surface state transitions between 1/2 O$^*$ and 3/4 O$^*$ regions, which we inspect further using $c_{sp=i}(U,pH=0)$ (refer to section \ref{sec:cvalue_pourbaix}) for all $i$ corresponding to the considered surface phases. Based on this uncertainty quantification measure, for each surface phase we depict the agreement within the BEEF-vdW ensemble of functionals in figure \ref{fig_Pd111}b. We notice a low degree of prediction confidence as to the lowest free energy surface phase at $\approx$0.72 V (vs. RHE) (pH=0) where the maximum c-value is $<0.3$. The 1/3 OH$^*$ surface has a relatively higher confidence value than the 1/4 O$^*$ surface suggesting that the clean surface transitions directly to the 1/3 OH$^*$ surface phase, an observation enabled by the defined quantity $c_{sp=i}(U,pH)$. Experimentally, Hara et al.\cite{hara2007preparation} ascribe broad anodic and cathodoic current peaks to OH$^*$ species around 0.7 V based on cyclic voltammetry in perchloric acid solution. We quantify the maximum observed c-value for 1/3 OH$^*$ between 0.6-0.7 V (vs. RHE) which lies close to the experimentally observed region within the prediction uncertainty. 
We observe comparatively large prediction confidence values for 1/3 O$^*$ and 1/2 O$^*$ covered surfaces between 0.8-1.3 V (vs. RHE), which is supported by Hara et al. who report an oxidation peak corresponding to oxygen adsorption at 1.02 V.

\subsection{Ir(111)}

\begin{figure}
	\centering
	\includegraphics[width=15cm,trim=1 1 20 20,clip]{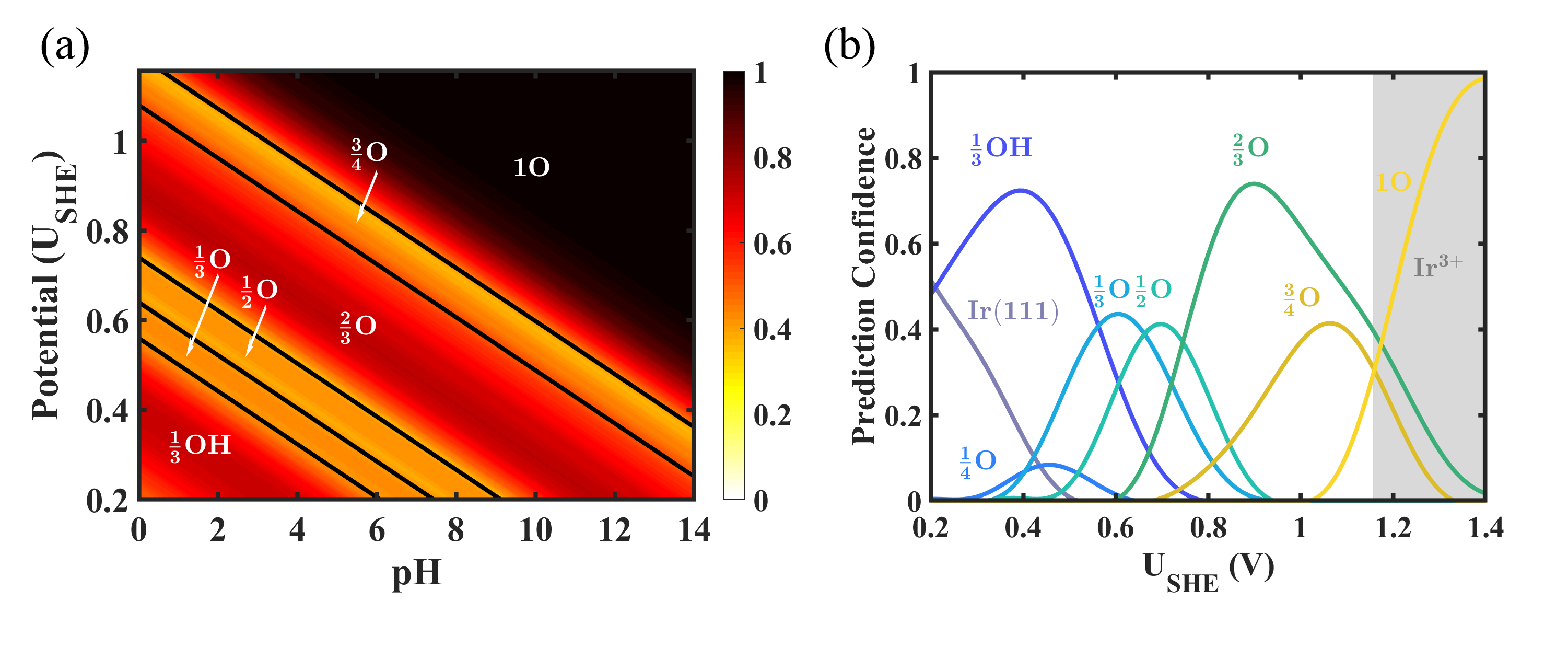}
	\caption{Figure (a) shows the surface Pourbaix diagram for Ir(111), where we observe oxidation to begin with 1/3 OH$^*$ in the range from 0.2 V to 0.55 V. We then see increase in oxidation until the dissolution potential at 1.16 V. Figure (b) shows each stable region assigned prediction confidence on the surface Pourbaix at pH=0 where we see a region with low confidence between 0.55 V and 0.7 V. In this region the surface may covered by 1/3 OH$^*$, 1/3 O$^*$, 1/2 O$^*$, or 2/3 O$^*$. The shaded gray region on the x-axis represents dissolution into Ir$^{3+}$.}
	\label{fig_Ir111}
\end{figure}

Iridium is a highly corrosion resistant platinum group metal and is similar to Pt although relatively less extensively studied\cite{cherevko2016oxygen}. 
Stable surface phases determined with the best-fit BEEF-vdW functional are shown with the bold black lines in the constructed surface Pourbaix diagram in figure \ref{fig_Ir111}a. Oxygen coverage begins with 1/3 OH$^*$ to potentials of $\approx$0.56 V (vs. RHE) at pH=0, consistent with prior DFT predictions that Ir is more oxophilic than Pt(111).\cite{krekelberg2004atomic,karlberg2006adsorption} The surface transitions to 1/3 O$^*$ at $\approx$0.56 V, followed by 1/2 O$^*$ at $\approx$0.64 V, 2/3 O$^*$ at $\approx$0.74 V, and 3/4 O$^*$ at $\approx$1.1 V (all vs. RHE), with dissolution at $\approx$1.16 V.\cite{lide1994crc} However, prediction confidence values corresponding to the 1/3 O$^*$, 1/2 O$^*$, and 3/4 O$^*$ surface phases are low (c-value $< 0.4$) relative to 1/3 OH$^*$ phase. To further probe the degree of agreement within the ensemble of functionals as to the lowest free energy phase especially close to the phase boundaries, we assess the uncertainty quantification metric $c_{sp=i}$ for all $i$ corresponding to the considered surface phases as shown in figure \ref{fig_Ir111}b. We predict with high confidence the 1/3 OH$^*$ adsorption phase approximately in the $0.2<U<0.55$ V (vs. RHE) range where the c-value is $>0.5$ with a maximum at $\approx$0.4 V. We observe that between $\approx0.55$ V  and $\approx0.75$ V (vs. RHE) a smooth transition from the 1/3 O$^*$ phase to the 1/2 O$^*$ phase. At higher potentials until dissolution, with a large degree of agreement between the functionals we predict the 2/3 O$^*$ surface phase to be the lowest in free energy. The 3/4 O$^*$ surface has a maximum c-value of nearly 0.4 at $\approx$1.0 V (vs. RHE), therefore, within DFT uncertainty we expect a short voltage window where this phase is stable just before dissolution. We note some uncertainty concerning the interpretation of peaks in cyclic voltammograms while comparing the findings from our analysis.  Cyclic voltammograms for Ir(111) in perchloric acid shows a broad peak between approximately 0.05 V to 0.35 V (vs. RHE).\cite{ganassin2017ph, pajkossy2005voltammetry,wan1999situ} In this range, Marc Koper suggests that the surface is covered with hydrogen but also tentatively suggests that the peak could represent a mixed transition phase from being OH-covered to being H-covered,\cite{koper2011blank} while Wan et al. have attributed this region to hydrogen adsorption.\cite{wan1999situ} We find that our prediction of a stronger binding OH$^*$ surface as compared to Pt(111) is closer in agreement with previous DFT calculations.\cite{karlberg2006adsorption} Similarly, regarding the interpretation of a pair of sharp cyclic voltammetry peaks at 0.92 V and 0.95 V in perchloric acid, Wan et al.\cite{wan1999situ} attributes it to initial stages of oxidation representing hydroxide adsorption, but this would depend on the interpretation of the peak between 0.05 V and 0.35 V since that would affect the length of the double-layer region.\cite{koper2011blank} With high confidence (c-value >0.7), we observe that the surface is O$^*$ covered at 0.9 V (vs. RHE) which is in agreement with DFT calculations\cite{karlberg2006adsorption}. We highlight that the defined confidence values provide a way to incorporate DFT uncertainty to interpret smooth surface phase transitions and aid a more integrated theory-experiment effort that is necessary for conclusive peak assignments.

\subsection{Rh(111)}

\begin{figure}
	\centering
	\includegraphics[width=15cm,trim=1 1 20 20,clip]{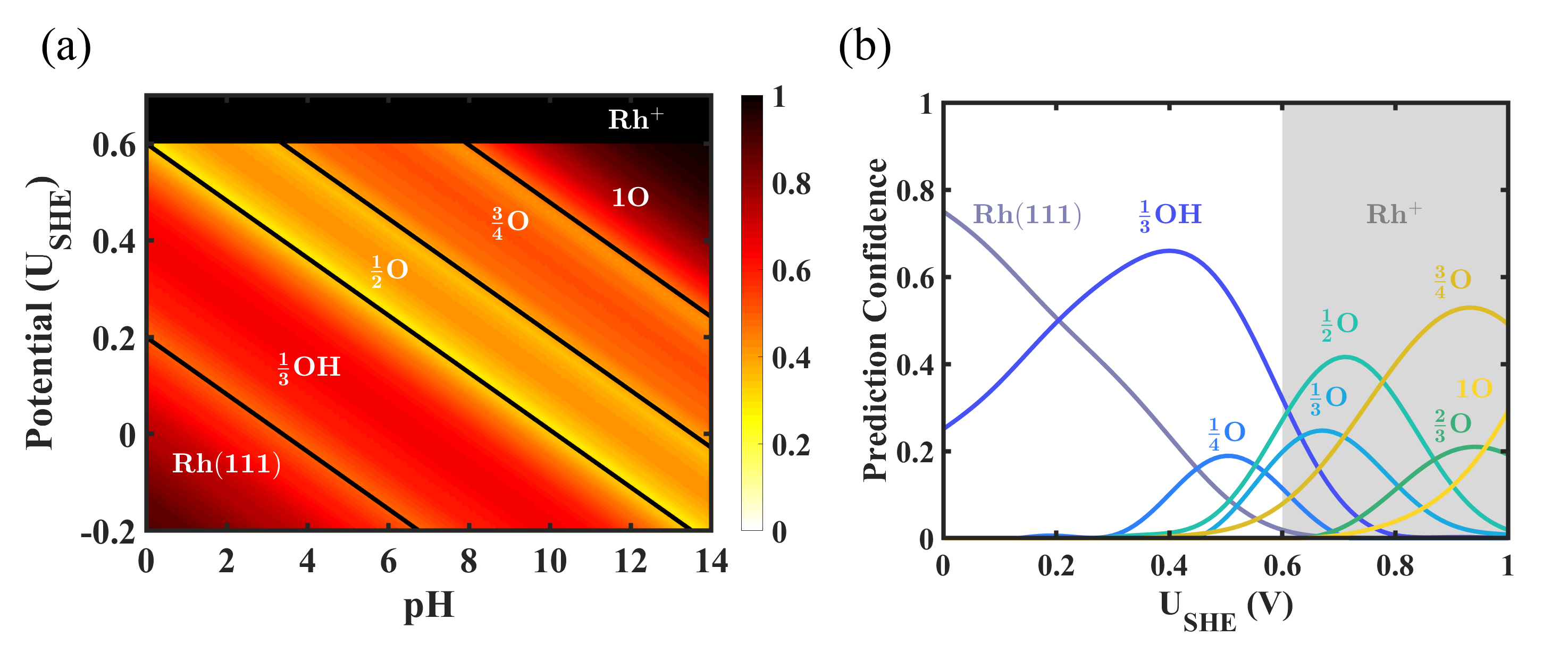}
	\caption{Figure (a) shows the surface Pourbaix diagram for Rh(111). Figure (b) describes the confidence for each surface state at pH=0 where dissolution from solid Rh to Rh$^{1+}$ is shown at 0.60 V by the shaded gray region. }
	\label{fig_Rh111}
\end{figure}

The Pt-group metal rhodium is less frequently studied since it is scarce and expensive. However, it is a versatile catalyst most known for use in the three-way-catalyst (TWC) for NO$_x$ reduction, and oxidation of CO and unburned hydrocarbons in automotive exhaust.\cite{heck2001automobile} We show the constructed surface Pourbaix diagram for the Rh(111) surface in figure \ref{fig_Rh111}a. At pH=0, we observe the 1/3 OH$^*$ surface phase to be stable between $\approx$0.2-0.6 V (vs. RHE). The surface Pourbaix diagram also shows that Rh(111) binds oxygen intermediates stronger than Pt(111) which is supported by prior DFT studies.\cite{karlberg2006adsorption,markovic2002surface} At pH=0, we show that dissolution is favored at 0.6 V (vs. RHE) of the crystalline Rh metal governed by the equilibrium relation $\mathrm{Rh^+ + e^- \rightleftharpoons Rh}$. Based on the best-fit BEEF-vdW functional, at a constant electrode potential with increasing pH, we observe that the 1/2 O$^*$ surface appears as the most stable phase followed by the 3/4 O$^*$ phase and subsequently the O$^*$ surface phase. However, we notice that the 1/2 O$^*$ phase is predicted to have a narrow stability region associated with a comparatively low (c-value < 0.4) degree of prediction confidence. 

We understand the relative agreement between the functionals as to the lowest free energy phase by constructing figure \ref{fig_Rh111}b based on $c_{sp=i}(U,pH)$ for all $i$ corresponding to the surfaces phases considered. We observe that 1/4 O$^*$, 1/3 O$^*$, and 2/3 O$^*$ surface phases are associated with low confidence values ($<0.3$) in accordance with the fact that these phases do no appear in figure \ref{fig_Rh111}a based on the best-fit BEEF-vdW functional. From figure \ref{fig_Rh111}b at pH=0 we predict the $\frac{1}{3}$OH$^*$ surface to be stable between $\approx$0.2 and $\approx$0.6 V (vs. RHE) with relatively high prediction confidence. Several experimental groups have studied cyclic voltammetry on Rh(111) in perchloric solutions to understand the sharp reversible peak visible at 0.64 V\cite{housmans2005co,clavilier1994detailed,wan1995situ}. The reversible nature of this peak has led it to be tentatively ascribed to OH adsorption\cite{housmans2005co,koper2011blank} although we predict the onset of 1/3 OH$^*$ phase at a lower potential. However, we cautiously speculate that a surface reconstruction to RhO$_2$ occurs based on the our prediction of the onset of OH$^*$ adsorption on RhO$_2$ at 0.66 V and the observed reversible butterfly peak at 0.64 V (refer to section 4 of the Supporting Information for additional details). This reconstruction could correspond to the broad peak at $\approx$0.3 V and also explain why the dissolution potential for Rh to Rh$^1+$ is experimentally observed\cite{lide1994crc} at 0.6 V while the cyclic voltammogram extends until $\approx$ 0.8 V.\cite{housmans2005co,koper2011blank}
However, further insight into the stability of the Rh oxide through a surface Pourbaix diagram is necessary to validate the hypothesis.  

\subsection{Ru(0001)}

\begin{figure}
	\centering
	\includegraphics[width=15cm,trim=1 1 20 20,clip]{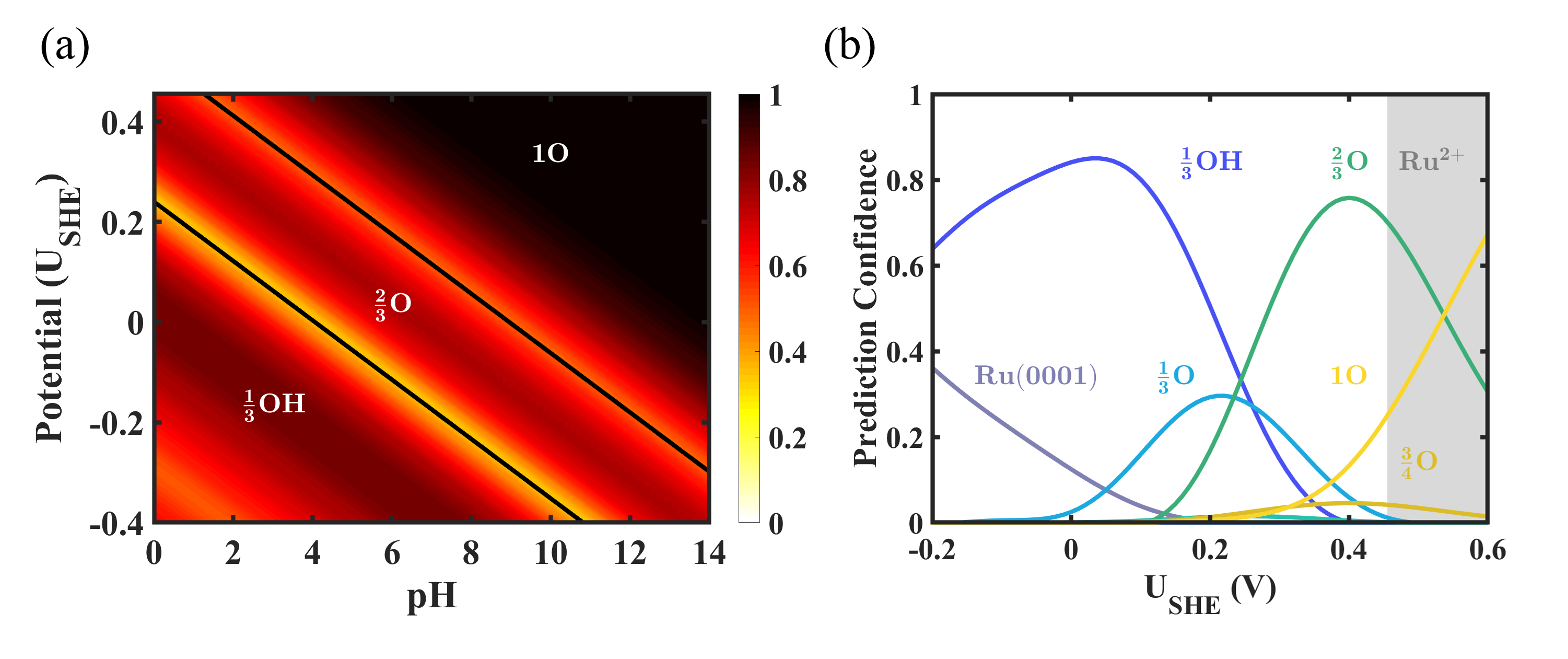}
	\caption{Figure (a) shows the surface Pourbaix diagram for Ru(0001) showing strong oxidation at low potentials. Figure (b) shows the confidence prediction for each surface state at pH of 0. The region on the x-axis shaded in gray shows the dissolution potential occurring at 0.46 V.}
	\label{fig_Ru0001}
\end{figure}

Ruthenium is used as a catalyst for nitrogen reduction\cite{jacobsen2000structure,aika1972activation} and is the basis for several bimetallic catalyst systems in electrocatalysts.\cite{hansen2002support,jackson2014climbing} We consider the widely studied Ru(0001) surface for our analysis. The surface Pourbaix diagram for the  Hexagonal close packed Ru(0001) surface is shown in figure \ref{fig_Ru0001}a with the surface transition from 1/3 OH$^*$ to 2/3 O$^*$ surface configuration at 0.24 V (vs. RHE) at a pH=0. At a constant pH, with increasing potential, we observe a transition from the 2/3 O$^*$ surface phase to the 1 O$^*$ phase. Ru metal binds oxygen strongly with lower onset potentials for adsorption of oxygenated species relative to other transition metals, which is in agreement with a previous DFT study by Taylor et al\cite{taylor2007first}. We highlight from figure \ref{fig_Ru0001}a that the stable surface states appearing in the Pourbaix diagram are predicted with a high level of confidence ($\approx$>0.5). We observe only the 1/3 OH$^*$ to 2/3 O$^*$ transition to have a comparatively low prediction confidence ($<0.37$). We assess the relative prediction confidence of all surface phases using $c_{sp=i}$ by constructing figure \ref{fig_Ru0001}b at pH=0. The maximum prediction confidence (c-value of 0.85) associated with the 1/3 OH$^*$ phase occurs near 0.03 V (vs. RHE). The dissolution potential of 0.46 V represents dissociation of crystalline Ru to Ru$^{2+}$ under standard conditions.\cite{lide1994crc} Our prediction of the OH$^*$ to O$^*$ phase transition at 0.24 V (vs. RHE) is close to the that reported by M. Koper (0.18 V) on considering the DFT prediction error ($\sigma_{OH}$=0.27, $\sigma_{O}$=0.09). While earlier computational work suggests a narrow double layer region,\cite{taylor2007first}, we do not observe a double layer region ($>-0.2$ V vs. RHE) with a high degree of agreement between functionals consistent with the interpretation of cyclic voltammograms\cite{zhou2007electrochemical,hoster2004catalytic,el2002new} by M. Koper, which highlights the usefulness of uncertainty incorporation. We rationalize the fact that cyclic voltammogram extends beyond 0.8 V while at standard conditions the dissolution of Ru to Ru$^{2+}$ is marked at 0.46 V based on a possible metal surface reconstruction to RuO$_2$, which is supported by a better agreement between our predicted onset of OH$^*$ adsorption on RuO$_2$ at 0.23 V and the observed reversible peak at 0.18 V. However, it remains to be confirmed that the OH$^*$ surface phase is the most stable in this potential regime through a Pourbaix diagram construction on RuO$_2$. 

Further improvements to the uncertainty incorporation method presented here can be made by accounting for configurational and vibrational entropy and accounting for uncertainties in zero point energy although we expect these effects to much smaller than the uncertainties accounted for in this paper.  For increased prediction accuracy of surface reactivity of catalysts, systematically accounting for a self-consistent loop between the stable state of the surface determined by surface Pourbaix diagrams and the reactivity determined by the limiting potential derived from free energy diagrams is essential. In addition, this work enables a precise determination of the stable catalyst surface state, which has important implications for identifying dominant reaction mechanisms and selectivity.

\section{Conclusions}
Surface Pourbaix diagrams are crucial in determining electrocatalytic reaction mechanisms and in corrosion processes.  In this work, we demonstrate a method to quantify uncertainty in density functional theory calculated surface Pourbaix diagrams. Based on this method, we define a confidence value (c-value) associated with each possible surface configuration at a given electrode potential, U, and pH. The approach involving the BEEF-vdW exchange correlation functional with built-in error estimation capabilities is much more computationally efficient relative to performing many calculations based on different functionals since this involves non-self-consistent calculations based on the converged density from one self-consistent calculation.
Using this method, we have constructed surface Pourbaix diagrams for each of Pt(111), Pt(100), Pd(111), Ir(111), Rh(111), and Ru(0001) surface and the associated confidence values for these surface phases. On Pt(111), our method captures the well-known uncertainty in the phase transition from OH$^*$ to O$^*$. We find good agreement between our predicted phase diagrams on Pt(100) and Pd(111) compared to cyclic voltammetry experiments for the onset of OH$^*$ and O$^*$ covered surfaces. Ir(111), Rh(111), and Ru(0001) all exhibit strong oxidation of 1/3 OH$^*$ which is supports prior theoretical studies. We also briefly consider oxidation on rutile metal-oxides as a possible explanation to connect DFT and experimental observations.  We suggest that reconstruction of the surface, particularly for Rh(111) and Ru(0001), could be important. An important implication from this work is the need to incorporate multiple surface phases in determining electrochemical reactions mechanisms and the associated activity due to finite uncertainty associated with predicting the stable state of the surface.

\section*{Acknowledgement}
O.V. gratefully acknowledges funding support in part from Volkswagen AG. D.K. and V.V gratefully acknowledge funding support from the National Science Foundation under award CBET-1554273.
\bibliographystyle{achemso}
\bibliography{refs}

\providecommand{\latin}[1]{#1}
\makeatletter
\providecommand{\doi}
  {\begingroup\let\do\@makeother\dospecials
  \catcode`\{=1 \catcode`\}=2 \doi@aux}
\providecommand{\doi@aux}[1]{\endgroup\texttt{#1}}
\makeatother
\providecommand*\mcitethebibliography{\thebibliography}
\csname @ifundefined\endcsname{endmcitethebibliography}
  {\let\endmcitethebibliography\endthebibliography}{}
\begin{mcitethebibliography}{61}
\providecommand*\natexlab[1]{#1}
\providecommand*\mciteSetBstSublistMode[1]{}
\providecommand*\mciteSetBstMaxWidthForm[2]{}
\providecommand*\mciteBstWouldAddEndPuncttrue
  {\def\EndOfBibitem{\unskip.}}
\providecommand*\mciteBstWouldAddEndPunctfalse
  {\let\EndOfBibitem\relax}
\providecommand*\mciteSetBstMidEndSepPunct[3]{}
\providecommand*\mciteSetBstSublistLabelBeginEnd[3]{}
\providecommand*\EndOfBibitem{}
\mciteSetBstSublistMode{f}
\mciteSetBstMaxWidthForm{subitem}{(\alph{mcitesubitemcount})}
\mciteSetBstSublistLabelBeginEnd
  {\mcitemaxwidthsubitemform\space}
  {\relax}
  {\relax}

\bibitem[Lewis and Nocera(2006)Lewis, and Nocera]{lewis2006powering}
Lewis,~N.~S.; Nocera,~D.~G. Powering the planet: Chemical challenges in solar
  energy utilization. \emph{PNAS} \textbf{2006}, \emph{103}, 15729--15735\relax
\mciteBstWouldAddEndPuncttrue
\mciteSetBstMidEndSepPunct{\mcitedefaultmidpunct}
{\mcitedefaultendpunct}{\mcitedefaultseppunct}\relax
\EndOfBibitem
\bibitem[Montoya \latin{et~al.}(2017)Montoya, Seitz, Chakthranont, Vojvodic,
  Jaramillo, and N{\o}rskov]{montoya2017materials}
Montoya,~J.~H.; Seitz,~L.~C.; Chakthranont,~P.; Vojvodic,~A.; Jaramillo,~T.~F.;
  N{\o}rskov,~J.~K. Materials for solar fuels and chemicals. \emph{Nat. Mater.}
  \textbf{2017}, \emph{16}, 70--81\relax
\mciteBstWouldAddEndPuncttrue
\mciteSetBstMidEndSepPunct{\mcitedefaultmidpunct}
{\mcitedefaultendpunct}{\mcitedefaultseppunct}\relax
\EndOfBibitem
\bibitem[Bockris and Khan(2013)Bockris, and Khan]{bockris2013surface}
Bockris,~J.~O.; Khan,~S.~U. \emph{Surface electrochemistry: a molecular level
  approach}; Springer Science \& Business Media, 2013\relax
\mciteBstWouldAddEndPuncttrue
\mciteSetBstMidEndSepPunct{\mcitedefaultmidpunct}
{\mcitedefaultendpunct}{\mcitedefaultseppunct}\relax
\EndOfBibitem
\bibitem[Bard \latin{et~al.}(1993)Bard, Abruna, Chidsey, Faulkner, Feldberg,
  Itaya, Majda, Melroy, and Murray]{bard1993electrode}
Bard,~A.~J.; Abruna,~H.~D.; Chidsey,~C.~E.; Faulkner,~L.~R.; Feldberg,~S.~W.;
  Itaya,~K.; Majda,~M.; Melroy,~O.; Murray,~R.~W. The electrode/electrolyte
  interface-a status report. \emph{J. Phys. Chem.} \textbf{1993}, \emph{97},
  7147--7173\relax
\mciteBstWouldAddEndPuncttrue
\mciteSetBstMidEndSepPunct{\mcitedefaultmidpunct}
{\mcitedefaultendpunct}{\mcitedefaultseppunct}\relax
\EndOfBibitem
\bibitem[Wakisaka \latin{et~al.}(2009)Wakisaka, Suzuki, Mitsui, Uchida, and
  Watanabe]{wakisaka2009identification}
Wakisaka,~M.; Suzuki,~H.; Mitsui,~S.; Uchida,~H.; Watanabe,~M. Identification
  and quantification of oxygen species adsorbed on Pt (111) single-crystal and
  polycrystalline Pt electrodes by photoelectron spectroscopy. \emph{Langmuir}
  \textbf{2009}, \emph{25}, 1897--1900\relax
\mciteBstWouldAddEndPuncttrue
\mciteSetBstMidEndSepPunct{\mcitedefaultmidpunct}
{\mcitedefaultendpunct}{\mcitedefaultseppunct}\relax
\EndOfBibitem
\bibitem[Casalongue \latin{et~al.}(2013)Casalongue, Kaya, Viswanathan, Miller,
  Friebel, Hansen, N{\o}rskov, Nilsson, and Ogasawara]{casalongue2013direct}
Casalongue,~H.~S.; Kaya,~S.; Viswanathan,~V.; Miller,~D.~J.; Friebel,~D.;
  Hansen,~H.~A.; N{\o}rskov,~J.~K.; Nilsson,~A.; Ogasawara,~H. Direct
  observation of the oxygenated species during oxygen reduction on a platinum
  fuel cell cathode. \emph{Nat. Commun.} \textbf{2013}, \emph{4}, 2817\relax
\mciteBstWouldAddEndPuncttrue
\mciteSetBstMidEndSepPunct{\mcitedefaultmidpunct}
{\mcitedefaultendpunct}{\mcitedefaultseppunct}\relax
\EndOfBibitem
\bibitem[Bondarenko \latin{et~al.}(2011)Bondarenko, Stephens, Hansen,
  P{\'e}rez-Alonso, Tripkovic, Johansson, Rossmeisl, N{\o}rskov, and
  Chorkendorff]{bondarenko2011pt}
Bondarenko,~A.~S.; Stephens,~I.~E.; Hansen,~H.~A.; P{\'e}rez-Alonso,~F.~J.;
  Tripkovic,~V.; Johansson,~T.~P.; Rossmeisl,~J.; N{\o}rskov,~J.~K.;
  Chorkendorff,~I. The Pt (111)/electrolyte interface under oxygen reduction
  reaction conditions: an electrochemical impedance spectroscopy study.
  \emph{Langmuir} \textbf{2011}, \emph{27}, 2058--2066\relax
\mciteBstWouldAddEndPuncttrue
\mciteSetBstMidEndSepPunct{\mcitedefaultmidpunct}
{\mcitedefaultendpunct}{\mcitedefaultseppunct}\relax
\EndOfBibitem
\bibitem[Hansen \latin{et~al.}(2008)Hansen, Rossmeisl, and
  N{\o}rskov]{hansen2008surface}
Hansen,~H.~A.; Rossmeisl,~J.; N{\o}rskov,~J.~K. Surface Pourbaix diagrams and
  oxygen reduction activity of Pt, Ag and Ni (111) surfaces studied by DFT.
  \emph{Phys. Chem. Chem. Phys.} \textbf{2008}, \emph{10}, 3722--3730\relax
\mciteBstWouldAddEndPuncttrue
\mciteSetBstMidEndSepPunct{\mcitedefaultmidpunct}
{\mcitedefaultendpunct}{\mcitedefaultseppunct}\relax
\EndOfBibitem
\bibitem[Tian \latin{et~al.}(2009)Tian, Jinnouchi, and
  Anderson]{tian2009potentials}
Tian,~F.; Jinnouchi,~R.; Anderson,~A.~B. How potentials of zero charge and
  potentials for water oxidation to OH (ads) on Pt (111) electrodes vary with
  coverage. \emph{J. Phys. Chem. C} \textbf{2009}, \emph{113},
  17484--17492\relax
\mciteBstWouldAddEndPuncttrue
\mciteSetBstMidEndSepPunct{\mcitedefaultmidpunct}
{\mcitedefaultendpunct}{\mcitedefaultseppunct}\relax
\EndOfBibitem
\bibitem[Jinnouchi and Anderson(2008)Jinnouchi, and
  Anderson]{jinnouchi2008aqueous}
Jinnouchi,~R.; Anderson,~A.~B. Aqueous and surface redox potentials from
  self-consistently determined Gibbs energies. \emph{J. Phys. Chem. C}
  \textbf{2008}, \emph{112}, 8747--8750\relax
\mciteBstWouldAddEndPuncttrue
\mciteSetBstMidEndSepPunct{\mcitedefaultmidpunct}
{\mcitedefaultendpunct}{\mcitedefaultseppunct}\relax
\EndOfBibitem
\bibitem[Wellendorff \latin{et~al.}(2012)Wellendorff, Lundgaard,
  M{\o}gelh{\o}j, Petzold, Landis, N{\o}rskov, Bligaard, and
  Jacobsen]{wellendorff2012density}
Wellendorff,~J.; Lundgaard,~K.~T.; M{\o}gelh{\o}j,~A.; Petzold,~V.;
  Landis,~D.~D.; N{\o}rskov,~J.~K.; Bligaard,~T.; Jacobsen,~K.~W. Density
  functionals for surface science: Exchange-correlation model development with
  Bayesian error estimation. \emph{Phys. Rev. B} \textbf{2012}, \emph{85},
  235149\relax
\mciteBstWouldAddEndPuncttrue
\mciteSetBstMidEndSepPunct{\mcitedefaultmidpunct}
{\mcitedefaultendpunct}{\mcitedefaultseppunct}\relax
\EndOfBibitem
\bibitem[Rossmeisl \latin{et~al.}(2006)Rossmeisl, N{\o}rskov, Taylor, Janik,
  and Neurock]{rossmeisl2006calculated}
Rossmeisl,~J.; N{\o}rskov,~J.~K.; Taylor,~C.~D.; Janik,~M.~J.; Neurock,~M.
  Calculated phase diagrams for the electrochemical oxidation and reduction of
  water over Pt (111). \emph{J. Phys. Chem. B} \textbf{2006}, \emph{110},
  21833--21839\relax
\mciteBstWouldAddEndPuncttrue
\mciteSetBstMidEndSepPunct{\mcitedefaultmidpunct}
{\mcitedefaultendpunct}{\mcitedefaultseppunct}\relax
\EndOfBibitem
\bibitem[Han \latin{et~al.}(2012)Han, Viswanathan, and Pitsch]{han2012first}
Han,~B.; Viswanathan,~V.; Pitsch,~H. First-principles based analysis of the
  electrocatalytic activity of the unreconstructed pt (100) surface for oxygen
  reduction reaction. \emph{J. Phys. Chem. C} \textbf{2012}, \emph{116},
  6174--6183\relax
\mciteBstWouldAddEndPuncttrue
\mciteSetBstMidEndSepPunct{\mcitedefaultmidpunct}
{\mcitedefaultendpunct}{\mcitedefaultseppunct}\relax
\EndOfBibitem
\bibitem[Hara \latin{et~al.}(2007)Hara, Linke, and
  Wandlowski]{hara2007preparation}
Hara,~M.; Linke,~U.; Wandlowski,~T. Preparation and electrochemical
  characterization of palladium single crystal electrodes in 0.1 MH 2 SO 4 and
  HClO 4: Part I. Low-index phases. \emph{Electrochim. Acta} \textbf{2007},
  \emph{52}, 5733--5748\relax
\mciteBstWouldAddEndPuncttrue
\mciteSetBstMidEndSepPunct{\mcitedefaultmidpunct}
{\mcitedefaultendpunct}{\mcitedefaultseppunct}\relax
\EndOfBibitem
\bibitem[Karlberg(2006)]{karlberg2006adsorption}
Karlberg,~G. Adsorption trends for water, hydroxyl, oxygen, and hydrogen on
  transition-metal and platinum-skin surfaces. \emph{Phys. Rev. B}
  \textbf{2006}, \emph{74}, 153414\relax
\mciteBstWouldAddEndPuncttrue
\mciteSetBstMidEndSepPunct{\mcitedefaultmidpunct}
{\mcitedefaultendpunct}{\mcitedefaultseppunct}\relax
\EndOfBibitem
\bibitem[Deshpande \latin{et~al.}(2016)Deshpande, Kitchin, and
  Viswanathan]{deshpande2016quantifying}
Deshpande,~S.; Kitchin,~J.~R.; Viswanathan,~V. Quantifying Uncertainty in
  Activity Volcano Relationships for Oxygen Reduction Reaction. \emph{ACS
  Catal.} \textbf{2016}, \emph{6}, 5251--5259\relax
\mciteBstWouldAddEndPuncttrue
\mciteSetBstMidEndSepPunct{\mcitedefaultmidpunct}
{\mcitedefaultendpunct}{\mcitedefaultseppunct}\relax
\EndOfBibitem
\bibitem[Krishnamurthy \latin{et~al.}(2018)Krishnamurthy, Sumaria, and
  Viswanathan]{krishnamurthy2018maximal}
Krishnamurthy,~D.; Sumaria,~V.; Viswanathan,~V. Maximal predictability approach
  for identifying the right descriptors for electrocatalytic reactions.
  \emph{J. Phys. Chem. Lett.} \textbf{2018}, \emph{9}, 588--595\relax
\mciteBstWouldAddEndPuncttrue
\mciteSetBstMidEndSepPunct{\mcitedefaultmidpunct}
{\mcitedefaultendpunct}{\mcitedefaultseppunct}\relax
\EndOfBibitem
\bibitem[Lide and Kehiaian(1994)Lide, and Kehiaian]{lide1994crc}
Lide,~D.~R.; Kehiaian,~H.~V. \emph{CRC handbook of thermophysical and
  thermochemical data}; Crc Press, 1994; Vol.~1\relax
\mciteBstWouldAddEndPuncttrue
\mciteSetBstMidEndSepPunct{\mcitedefaultmidpunct}
{\mcitedefaultendpunct}{\mcitedefaultseppunct}\relax
\EndOfBibitem
\bibitem[Perdew \latin{et~al.}(1996)Perdew, Burke, and
  Ernzerhof]{perdew1996generalized}
Perdew,~J.~P.; Burke,~K.; Ernzerhof,~M. Generalized gradient approximation made
  simple. \emph{Phys. Rev. Lett.} \textbf{1996}, \emph{77}, 3865\relax
\mciteBstWouldAddEndPuncttrue
\mciteSetBstMidEndSepPunct{\mcitedefaultmidpunct}
{\mcitedefaultendpunct}{\mcitedefaultseppunct}\relax
\EndOfBibitem
\bibitem[Lee \latin{et~al.}(2010)Lee, Murray, Kong, Lundqvist, and
  Langreth]{lee2010higher}
Lee,~K.; Murray,~{\'E}.~D.; Kong,~L.; Lundqvist,~B.~I.; Langreth,~D.~C.
  Higher-accuracy van der Waals density functional. \emph{Phys. Rev. B}
  \textbf{2010}, \emph{82}, 081101\relax
\mciteBstWouldAddEndPuncttrue
\mciteSetBstMidEndSepPunct{\mcitedefaultmidpunct}
{\mcitedefaultendpunct}{\mcitedefaultseppunct}\relax
\EndOfBibitem
\bibitem[Sumaria \latin{et~al.}(2018)Sumaria, Krishnamurthy, and
  Viswanathan]{sumaria2018quantifying}
Sumaria,~V.; Krishnamurthy,~D.; Viswanathan,~V. Quantifying Confidence in DFT
  Predicted Surface Pourbaix Diagrams and Associated Reaction Pathways for
  Chlorine Evolution. \emph{arXiv:1804.02766 [cond-mat.mtrl-sci]}
  \textbf{2018}, \relax
\mciteBstWouldAddEndPunctfalse
\mciteSetBstMidEndSepPunct{\mcitedefaultmidpunct}
{}{\mcitedefaultseppunct}\relax
\EndOfBibitem
\bibitem[Houchins and Viswanathan(2017)Houchins, and
  Viswanathan]{houchins2017quantifying}
Houchins,~G.; Viswanathan,~V. Quantifying confidence in density functional
  theory predictions of magnetic ground states. \emph{Phys. Rev. B}
  \textbf{2017}, \emph{96}, 134426\relax
\mciteBstWouldAddEndPuncttrue
\mciteSetBstMidEndSepPunct{\mcitedefaultmidpunct}
{\mcitedefaultendpunct}{\mcitedefaultseppunct}\relax
\EndOfBibitem
\bibitem[Garcia-Araez \latin{et~al.}(2009)Garcia-Araez, Climent, and
  Feliu]{garcia2009potential}
Garcia-Araez,~N.; Climent,~V.; Feliu,~J. Potential-dependent water orientation
  on Pt (111), Pt (100), and Pt (110), as inferred from laser-pulsed
  experiments. Electrostatic and chemical effects. \emph{J. Phys. Chem. C}
  \textbf{2009}, \emph{113}, 9290--9304\relax
\mciteBstWouldAddEndPuncttrue
\mciteSetBstMidEndSepPunct{\mcitedefaultmidpunct}
{\mcitedefaultendpunct}{\mcitedefaultseppunct}\relax
\EndOfBibitem
\bibitem[Ogasawara \latin{et~al.}(2002)Ogasawara, Brena, Nordlund, Nyberg,
  Pelmenschikov, Pettersson, and Nilsson]{ogasawara2002structure}
Ogasawara,~H.; Brena,~B.; Nordlund,~D.; Nyberg,~M.; Pelmenschikov,~A.;
  Pettersson,~L.; Nilsson,~A. Structure and bonding of water on Pt (111).
  \emph{Phys. Rev. Lett.} \textbf{2002}, \emph{89}, 276102\relax
\mciteBstWouldAddEndPuncttrue
\mciteSetBstMidEndSepPunct{\mcitedefaultmidpunct}
{\mcitedefaultendpunct}{\mcitedefaultseppunct}\relax
\EndOfBibitem
\bibitem[Viswanathan \latin{et~al.}(2012)Viswanathan, Hansen, Rossmeisl,
  Jaramillo, Pitsch, and N{\o}rskov]{viswanathan2012simulating}
Viswanathan,~V.; Hansen,~H.~A.; Rossmeisl,~J.; Jaramillo,~T.~F.; Pitsch,~H.;
  N{\o}rskov,~J.~K. Simulating linear sweep voltammetry from first-principles:
  application to electrochemical oxidation of water on Pt (111) and Pt3Ni
  (111). \emph{J. Phys. Chem. C} \textbf{2012}, \emph{116}, 4698--4704\relax
\mciteBstWouldAddEndPuncttrue
\mciteSetBstMidEndSepPunct{\mcitedefaultmidpunct}
{\mcitedefaultendpunct}{\mcitedefaultseppunct}\relax
\EndOfBibitem
\bibitem[Rossmeisl \latin{et~al.}(2009)Rossmeisl, Karlberg, Jaramillo, and
  N{\o}rskov]{rossmeisl2009steady}
Rossmeisl,~J.; Karlberg,~G.~S.; Jaramillo,~T.; N{\o}rskov,~J.~K. Steady state
  oxygen reduction and cyclic voltammetry. \emph{Faraday Discuss.}
  \textbf{2009}, \emph{140}, 337--346\relax
\mciteBstWouldAddEndPuncttrue
\mciteSetBstMidEndSepPunct{\mcitedefaultmidpunct}
{\mcitedefaultendpunct}{\mcitedefaultseppunct}\relax
\EndOfBibitem
\bibitem[Taylor \latin{et~al.}(2007)Taylor, Kelly, and
  Neurock]{taylor2007first}
Taylor,~C.~D.; Kelly,~R.~G.; Neurock,~M. First-principles prediction of
  equilibrium potentials for water activation by a series of metals. \emph{J.
  Electrochem. Soc.} \textbf{2007}, \emph{154}, F217--F221\relax
\mciteBstWouldAddEndPuncttrue
\mciteSetBstMidEndSepPunct{\mcitedefaultmidpunct}
{\mcitedefaultendpunct}{\mcitedefaultseppunct}\relax
\EndOfBibitem
\bibitem[Wang \latin{et~al.}(2004)Wang, Markovic, and Adzic]{wang2004kinetic}
Wang,~J.; Markovic,~N.; Adzic,~R. Kinetic analysis of oxygen reduction on Pt
  (111) in acid solutions: intrinsic kinetic parameters and anion adsorption
  effects. \emph{J. Phys. Chem. B} \textbf{2004}, \emph{108}, 4127--4133\relax
\mciteBstWouldAddEndPuncttrue
\mciteSetBstMidEndSepPunct{\mcitedefaultmidpunct}
{\mcitedefaultendpunct}{\mcitedefaultseppunct}\relax
\EndOfBibitem
\bibitem[Koper(2011)]{koper2011blank}
Koper,~M. Blank voltammetry of hexagonal surfaces of Pt-group metal electrodes:
  comparison to density functional theory calculations and ultra-high vacuum
  experiments on water dissociation. \emph{Electrochim. Acta} \textbf{2011},
  \emph{56}, 10645--10651\relax
\mciteBstWouldAddEndPuncttrue
\mciteSetBstMidEndSepPunct{\mcitedefaultmidpunct}
{\mcitedefaultendpunct}{\mcitedefaultseppunct}\relax
\EndOfBibitem
\bibitem[Garcia-Araez \latin{et~al.}(2008)Garcia-Araez, Climent, and
  Feliu]{garcia2008determination}
Garcia-Araez,~N.; Climent,~V.; Feliu,~J.~M. Determination of the entropy of
  formation of the Pt (111)|perchloric acid solution interface. Estimation of
  the entropy of adsorbed hydrogen and OH species. \emph{J. Solid State
  Electrochem.} \textbf{2008}, \emph{12}, 387--398\relax
\mciteBstWouldAddEndPuncttrue
\mciteSetBstMidEndSepPunct{\mcitedefaultmidpunct}
{\mcitedefaultendpunct}{\mcitedefaultseppunct}\relax
\EndOfBibitem
\bibitem[Garcia-Araez \latin{et~al.}(2006)Garcia-Araez, Climent, Herrero,
  Feliu, and Lipkowski]{garcia2006thermodynamic}
Garcia-Araez,~N.; Climent,~V.; Herrero,~E.; Feliu,~J.~M.; Lipkowski,~J.
  Thermodynamic approach to the double layer capacity of a Pt (111) electrode
  in perchloric acid solutions. \emph{Electrochim. Acta} \textbf{2006},
  \emph{51}, 3787--3793\relax
\mciteBstWouldAddEndPuncttrue
\mciteSetBstMidEndSepPunct{\mcitedefaultmidpunct}
{\mcitedefaultendpunct}{\mcitedefaultseppunct}\relax
\EndOfBibitem
\bibitem[Stamenkovic \latin{et~al.}(2001)Stamenkovic, Markovic, and
  Ross~Jr]{stamenkovic2001structure}
Stamenkovic,~V.; Markovic,~N.; Ross~Jr,~P. Structure-relationships in
  electrocatalysis: oxygen reduction and hydrogen oxidation reactions on Pt
  (111) and Pt (100) in solutions containing chloride ions. \emph{J.
  Electroanal. Chem.} \textbf{2001}, \emph{500}, 44--51\relax
\mciteBstWouldAddEndPuncttrue
\mciteSetBstMidEndSepPunct{\mcitedefaultmidpunct}
{\mcitedefaultendpunct}{\mcitedefaultseppunct}\relax
\EndOfBibitem
\bibitem[Climent \latin{et~al.}(2006)Climent, G{\'o}mez, Orts, and
  Feliu]{climent2006thermodynamic}
Climent,~V.; G{\'o}mez,~R.; Orts,~J.~M.; Feliu,~J.~M. Thermodynamic analysis of
  the temperature dependence of OH adsorption on Pt (111) and Pt (100)
  electrodes in acidic media in the absence of specific anion adsorption.
  \emph{J. Phys. Chem. B} \textbf{2006}, \emph{110}, 11344--11351\relax
\mciteBstWouldAddEndPuncttrue
\mciteSetBstMidEndSepPunct{\mcitedefaultmidpunct}
{\mcitedefaultendpunct}{\mcitedefaultseppunct}\relax
\EndOfBibitem
\bibitem[G{\'o}mez \latin{et~al.}(2004)G{\'o}mez, Orts, {\'A}lvarez-Ruiz, and
  Feliu]{gomez2004effect}
G{\'o}mez,~R.; Orts,~J.~M.; {\'A}lvarez-Ruiz,~B.; Feliu,~J.~M. Effect of
  temperature on hydrogen adsorption on Pt (111), Pt (110), and Pt (100)
  electrodes in 0.1 M HClO4. \emph{J. Phys. Chem. B} \textbf{2004}, \emph{108},
  228--238\relax
\mciteBstWouldAddEndPuncttrue
\mciteSetBstMidEndSepPunct{\mcitedefaultmidpunct}
{\mcitedefaultendpunct}{\mcitedefaultseppunct}\relax
\EndOfBibitem
\bibitem[Duan and Wang(2013)Duan, and Wang]{duan2013comparison}
Duan,~Z.; Wang,~G. Comparison of reaction energetics for oxygen reduction
  reactions on Pt (100), Pt (111), Pt/Ni (100), and Pt/Ni (111) surfaces: a
  first-principles study. \emph{J. Phys. Chem. C} \textbf{2013}, \emph{117},
  6284--6292\relax
\mciteBstWouldAddEndPuncttrue
\mciteSetBstMidEndSepPunct{\mcitedefaultmidpunct}
{\mcitedefaultendpunct}{\mcitedefaultseppunct}\relax
\EndOfBibitem
\bibitem[Song \latin{et~al.}(2005)Song, Kim, Connor, Somorjai, and
  Yang]{song2005pt}
Song,~H.; Kim,~F.; Connor,~S.; Somorjai,~G.~A.; Yang,~P. Pt nanocrystals: shape
  control and langmuir- blodgett monolayer formation. \emph{J. Phys. Chem. B}
  \textbf{2005}, \emph{109}, 188--193\relax
\mciteBstWouldAddEndPuncttrue
\mciteSetBstMidEndSepPunct{\mcitedefaultmidpunct}
{\mcitedefaultendpunct}{\mcitedefaultseppunct}\relax
\EndOfBibitem
\bibitem[Wang \latin{et~al.}(2008)Wang, Daimon, Onodera, Koda, and
  Sun]{wang2008general}
Wang,~C.; Daimon,~H.; Onodera,~T.; Koda,~T.; Sun,~S. A General Approach to the
  Size-and Shape-Controlled Synthesis of Platinum Nanoparticles and Their
  Catalytic Reduction of Oxygen. \emph{Angew. Chem. Int. Ed.} \textbf{2008},
  \emph{47}, 3588--3591\relax
\mciteBstWouldAddEndPuncttrue
\mciteSetBstMidEndSepPunct{\mcitedefaultmidpunct}
{\mcitedefaultendpunct}{\mcitedefaultseppunct}\relax
\EndOfBibitem
\bibitem[Karlberg \latin{et~al.}(2007)Karlberg, Jaramillo, Skulason, Rossmeisl,
  Bligaard, and N{\o}rskov]{karlberg2007cyclic}
Karlberg,~G.; Jaramillo,~T.; Skulason,~E.; Rossmeisl,~J.; Bligaard,~T.;
  N{\o}rskov,~J.~K. Cyclic voltammograms for H on Pt (111) and Pt (100) from
  first principles. \emph{Phys. Rev. Lett.} \textbf{2007}, \emph{99},
  126101\relax
\mciteBstWouldAddEndPuncttrue
\mciteSetBstMidEndSepPunct{\mcitedefaultmidpunct}
{\mcitedefaultendpunct}{\mcitedefaultseppunct}\relax
\EndOfBibitem
\bibitem[Shao \latin{et~al.}(2006)Shao, Sasaki, and Adzic]{shao2006pd}
Shao,~M.-H.; Sasaki,~K.; Adzic,~R.~R. Pd- Fe nanoparticles as electrocatalysts
  for oxygen reduction. \emph{J. Am. Chem. Soc.} \textbf{2006}, \emph{128},
  3526--3527\relax
\mciteBstWouldAddEndPuncttrue
\mciteSetBstMidEndSepPunct{\mcitedefaultmidpunct}
{\mcitedefaultendpunct}{\mcitedefaultseppunct}\relax
\EndOfBibitem
\bibitem[Fern{\'a}ndez \latin{et~al.}(2005)Fern{\'a}ndez, Raghuveer, Manthiram,
  and Bard]{fernandez2005pd}
Fern{\'a}ndez,~J.~L.; Raghuveer,~V.; Manthiram,~A.; Bard,~A.~J. Pd- Ti and Pd-
  Co- Au electrocatalysts as a replacement for platinum for oxygen reduction in
  proton exchange membrane fuel cells. \emph{J. Am. Chem. Soc.} \textbf{2005},
  \emph{127}, 13100--13101\relax
\mciteBstWouldAddEndPuncttrue
\mciteSetBstMidEndSepPunct{\mcitedefaultmidpunct}
{\mcitedefaultendpunct}{\mcitedefaultseppunct}\relax
\EndOfBibitem
\bibitem[Fern{\'a}ndez \latin{et~al.}(2005)Fern{\'a}ndez, Walsh, and
  Bard]{fernandez2005thermodynamic}
Fern{\'a}ndez,~J.~L.; Walsh,~D.~A.; Bard,~A.~J. Thermodynamic guidelines for
  the design of bimetallic catalysts for oxygen electroreduction and rapid
  screening by scanning electrochemical microscopy. M- Co (M: Pd, Ag, Au).
  \emph{J. Am. Chem. Soc.} \textbf{2005}, \emph{127}, 357--365\relax
\mciteBstWouldAddEndPuncttrue
\mciteSetBstMidEndSepPunct{\mcitedefaultmidpunct}
{\mcitedefaultendpunct}{\mcitedefaultseppunct}\relax
\EndOfBibitem
\bibitem[Savadogo \latin{et~al.}(2004)Savadogo, Lee, Oishi, Mitsushima, Kamiya,
  and Ota]{savadogo2004new}
Savadogo,~O.; Lee,~K.; Oishi,~K.; Mitsushima,~S.; Kamiya,~N.; Ota,~K.-I. New
  palladium alloys catalyst for the oxygen reduction reaction in an acid
  medium. \emph{Electrochem. Commun.} \textbf{2004}, \emph{6}, 105--109\relax
\mciteBstWouldAddEndPuncttrue
\mciteSetBstMidEndSepPunct{\mcitedefaultmidpunct}
{\mcitedefaultendpunct}{\mcitedefaultseppunct}\relax
\EndOfBibitem
\bibitem[Mitsui \latin{et~al.}(2002)Mitsui, Rose, Fomin, Ogletree, and
  Salmeron]{mitsui2002water}
Mitsui,~T.; Rose,~M.; Fomin,~E.; Ogletree,~D.~F.; Salmeron,~M. Water diffusion
  and clustering on Pd (111). \emph{Science} \textbf{2002}, \emph{297},
  1850--1852\relax
\mciteBstWouldAddEndPuncttrue
\mciteSetBstMidEndSepPunct{\mcitedefaultmidpunct}
{\mcitedefaultendpunct}{\mcitedefaultseppunct}\relax
\EndOfBibitem
\bibitem[Cherevko \latin{et~al.}(2016)Cherevko, Geiger, Kasian, Kulyk, Grote,
  Savan, Shrestha, Merzlikin, Breitbach, and Ludwig]{cherevko2016oxygen}
Cherevko,~S.; Geiger,~S.; Kasian,~O.; Kulyk,~N.; Grote,~J.-P.; Savan,~A.;
  Shrestha,~B.~R.; Merzlikin,~S.; Breitbach,~B.; Ludwig,~A. Oxygen and hydrogen
  evolution reactions on Ru, RuO 2, Ir, and IrO 2 thin film electrodes in
  acidic and alkaline electrolytes: a comparative study on activity and
  stability. \emph{Catal. Today} \textbf{2016}, \emph{262}, 170--180\relax
\mciteBstWouldAddEndPuncttrue
\mciteSetBstMidEndSepPunct{\mcitedefaultmidpunct}
{\mcitedefaultendpunct}{\mcitedefaultseppunct}\relax
\EndOfBibitem
\bibitem[Krekelberg \latin{et~al.}(2004)Krekelberg, Greeley, and
  Mavrikakis]{krekelberg2004atomic}
Krekelberg,~W.~P.; Greeley,~J.; Mavrikakis,~M. Atomic and molecular adsorption
  on Ir (111). \emph{J. Phys. Chem. B} \textbf{2004}, \emph{108},
  987--994\relax
\mciteBstWouldAddEndPuncttrue
\mciteSetBstMidEndSepPunct{\mcitedefaultmidpunct}
{\mcitedefaultendpunct}{\mcitedefaultseppunct}\relax
\EndOfBibitem
\bibitem[Ganassin \latin{et~al.}(2017)Ganassin, Sebasti{\'a}n, Climent,
  Schuhmann, Bandarenka, and Feliu]{ganassin2017ph}
Ganassin,~A.; Sebasti{\'a}n,~P.; Climent,~V.; Schuhmann,~W.; Bandarenka,~A.~S.;
  Feliu,~J. On the pH Dependence of the Potential of Maximum Entropy of Ir
  (111) Electrodes. \emph{Sci. Rep.} \textbf{2017}, \emph{7}\relax
\mciteBstWouldAddEndPuncttrue
\mciteSetBstMidEndSepPunct{\mcitedefaultmidpunct}
{\mcitedefaultendpunct}{\mcitedefaultseppunct}\relax
\EndOfBibitem
\bibitem[Pajkossy \latin{et~al.}(2005)Pajkossy, Kibler, and
  Kolb]{pajkossy2005voltammetry}
Pajkossy,~T.; Kibler,~L.; Kolb,~D. Voltammetry and impedance measurements of Ir
  (111) electrodes in aqueous solutions. \emph{J. Electroanal. Chem.}
  \textbf{2005}, \emph{582}, 69--75\relax
\mciteBstWouldAddEndPuncttrue
\mciteSetBstMidEndSepPunct{\mcitedefaultmidpunct}
{\mcitedefaultendpunct}{\mcitedefaultseppunct}\relax
\EndOfBibitem
\bibitem[Wan \latin{et~al.}(1999)Wan, Hara, Inukai, Itaya, and
  et~al]{wan1999situ}
Wan,~L.-J.; Hara,~M.; Inukai,~J.; Itaya,~K.; et~al, In situ scanning tunneling
  microscopy of well-defined Ir (111) surface: high-resolution imaging of
  adsorbed sulfate. \emph{J. Phys. Chem. B} \textbf{1999}, \emph{103},
  6978--6983\relax
\mciteBstWouldAddEndPuncttrue
\mciteSetBstMidEndSepPunct{\mcitedefaultmidpunct}
{\mcitedefaultendpunct}{\mcitedefaultseppunct}\relax
\EndOfBibitem
\bibitem[Heck and Farrauto(2001)Heck, and Farrauto]{heck2001automobile}
Heck,~R.~M.; Farrauto,~R.~J. Automobile exhaust catalysts. \emph{Appl. Catal.,
  A} \textbf{2001}, \emph{221}, 443--457\relax
\mciteBstWouldAddEndPuncttrue
\mciteSetBstMidEndSepPunct{\mcitedefaultmidpunct}
{\mcitedefaultendpunct}{\mcitedefaultseppunct}\relax
\EndOfBibitem
\bibitem[Markovi{\'c} and Ross(2002)Markovi{\'c}, and
  Ross]{markovic2002surface}
Markovi{\'c},~N.; Ross,~P.~N. Surface science studies of model fuel cell
  electrocatalysts. \emph{Surf. Sci. Rep.} \textbf{2002}, \emph{45},
  117--229\relax
\mciteBstWouldAddEndPuncttrue
\mciteSetBstMidEndSepPunct{\mcitedefaultmidpunct}
{\mcitedefaultendpunct}{\mcitedefaultseppunct}\relax
\EndOfBibitem
\bibitem[Housmans and Koper(2005)Housmans, and Koper]{housmans2005co}
Housmans,~T.; Koper,~M. CO oxidation on stepped Rh [n (1 1 1)$\times$(1 1 1)]
  single crystal electrodes: Anion effects on CO surface mobility.
  \emph{Electrochem. Commun.} \textbf{2005}, \emph{7}, 581--588\relax
\mciteBstWouldAddEndPuncttrue
\mciteSetBstMidEndSepPunct{\mcitedefaultmidpunct}
{\mcitedefaultendpunct}{\mcitedefaultseppunct}\relax
\EndOfBibitem
\bibitem[Clavilier \latin{et~al.}(1994)Clavilier, Wasberg, Petit, and
  Klein]{clavilier1994detailed}
Clavilier,~J.; Wasberg,~M.; Petit,~M.; Klein,~L. Detailed analysis of the
  voltammetry of Rh (111) in perchloric acid solution. \emph{J. Electroanal.
  Chem.} \textbf{1994}, \emph{374}, 123--131\relax
\mciteBstWouldAddEndPuncttrue
\mciteSetBstMidEndSepPunct{\mcitedefaultmidpunct}
{\mcitedefaultendpunct}{\mcitedefaultseppunct}\relax
\EndOfBibitem
\bibitem[Wan \latin{et~al.}(1995)Wan, Yau, Swain, and Itaya]{wan1995situ}
Wan,~L.-J.; Yau,~S.-L.; Swain,~G.~M.; Itaya,~K. In-situ scanning tunneling
  microscopy of well-ordered Rh (111) electrodes. \emph{J. Electroanal. Chem.}
  \textbf{1995}, \emph{381}, 105--111\relax
\mciteBstWouldAddEndPuncttrue
\mciteSetBstMidEndSepPunct{\mcitedefaultmidpunct}
{\mcitedefaultendpunct}{\mcitedefaultseppunct}\relax
\EndOfBibitem
\bibitem[Jacobsen \latin{et~al.}(2000)Jacobsen, Dahl, Hansen, T{\"o}rnqvist,
  Jensen, Tops{\o}e, Prip, M{\o}enshaug, and
  Chorkendorff]{jacobsen2000structure}
Jacobsen,~C.~J.; Dahl,~S.; Hansen,~P.~L.; T{\"o}rnqvist,~E.; Jensen,~L.;
  Tops{\o}e,~H.; Prip,~D.~V.; M{\o}enshaug,~P.~B.; Chorkendorff,~I. Structure
  sensitivity of supported ruthenium catalysts for ammonia synthesis. \emph{J.
  Mol. Catal. A: Chem.} \textbf{2000}, \emph{163}, 19--26\relax
\mciteBstWouldAddEndPuncttrue
\mciteSetBstMidEndSepPunct{\mcitedefaultmidpunct}
{\mcitedefaultendpunct}{\mcitedefaultseppunct}\relax
\EndOfBibitem
\bibitem[Aika \latin{et~al.}(1972)Aika, Hori, and Ozaki]{aika1972activation}
Aika,~K.-i.; Hori,~H.; Ozaki,~A. Activation of nitrogen by alkali metal
  promoted transition metal I. Ammonia synthesis over ruthenium promoted by
  alkali metal. \emph{J. Catal.} \textbf{1972}, \emph{27}, 424--431\relax
\mciteBstWouldAddEndPuncttrue
\mciteSetBstMidEndSepPunct{\mcitedefaultmidpunct}
{\mcitedefaultendpunct}{\mcitedefaultseppunct}\relax
\EndOfBibitem
\bibitem[Hansen \latin{et~al.}(2002)Hansen, Hansen, Dahl, and
  Jacobsen]{hansen2002support}
Hansen,~T.~W.; Hansen,~P.~L.; Dahl,~S.; Jacobsen,~C.~J. Support effect and
  active sites on promoted ruthenium catalysts for ammonia synthesis.
  \emph{Catal. Lett.} \textbf{2002}, \emph{84}, 7--12\relax
\mciteBstWouldAddEndPuncttrue
\mciteSetBstMidEndSepPunct{\mcitedefaultmidpunct}
{\mcitedefaultendpunct}{\mcitedefaultseppunct}\relax
\EndOfBibitem
\bibitem[Jackson \latin{et~al.}(2014)Jackson, Viswanathan, Forman, Larsen,
  N{\o}rskov, and Jaramillo]{jackson2014climbing}
Jackson,~A.; Viswanathan,~V.; Forman,~A.~J.; Larsen,~A.~H.; N{\o}rskov,~J.~K.;
  Jaramillo,~T.~F. Climbing the activity volcano: core--shell Ru at Pt
  electrocatalysts for oxygen reduction. \emph{ChemElectroChem} \textbf{2014},
  \emph{1}, 67--71\relax
\mciteBstWouldAddEndPuncttrue
\mciteSetBstMidEndSepPunct{\mcitedefaultmidpunct}
{\mcitedefaultendpunct}{\mcitedefaultseppunct}\relax
\EndOfBibitem
\bibitem[Zhou \latin{et~al.}(2007)Zhou, Lewera, Bagus, and
  Wieckowski]{zhou2007electrochemical}
Zhou,~W.-P.; Lewera,~A.; Bagus,~P.~S.; Wieckowski,~A. Electrochemical and
  electronic properties of platinum deposits on Ru (0001): Combined XPS and
  cyclic voltammetric study. \emph{J. Phys. Chem. C} \textbf{2007}, \emph{111},
  13490--13496\relax
\mciteBstWouldAddEndPuncttrue
\mciteSetBstMidEndSepPunct{\mcitedefaultmidpunct}
{\mcitedefaultendpunct}{\mcitedefaultseppunct}\relax
\EndOfBibitem
\bibitem[Hoster \latin{et~al.}(2004)Hoster, Richter, and
  Behm]{hoster2004catalytic}
Hoster,~H.; Richter,~B.; Behm,~R. Catalytic influence of Pt monolayer islands
  on the hydrogen electrochemistry of Ru (0001) studied by ultrahigh vacuum
  scanning tunneling microscopy and cyclic voltammetry. \emph{J. Phys. Chem. B}
  \textbf{2004}, \emph{108}, 14780--14788\relax
\mciteBstWouldAddEndPuncttrue
\mciteSetBstMidEndSepPunct{\mcitedefaultmidpunct}
{\mcitedefaultendpunct}{\mcitedefaultseppunct}\relax
\EndOfBibitem
\bibitem[El-Aziz and Kibler(2002)El-Aziz, and Kibler]{el2002new}
El-Aziz,~A.; Kibler,~L. New information about the electrochemical behaviour of
  Ru (0 0 0 1) in perchloric acid solutions. \emph{Electrochem. Commun.}
  \textbf{2002}, \emph{4}, 866--870\relax
\mciteBstWouldAddEndPuncttrue
\mciteSetBstMidEndSepPunct{\mcitedefaultmidpunct}
{\mcitedefaultendpunct}{\mcitedefaultseppunct}\relax
\EndOfBibitem
\end{mcitethebibliography}
\includepdf[pages=1-17]{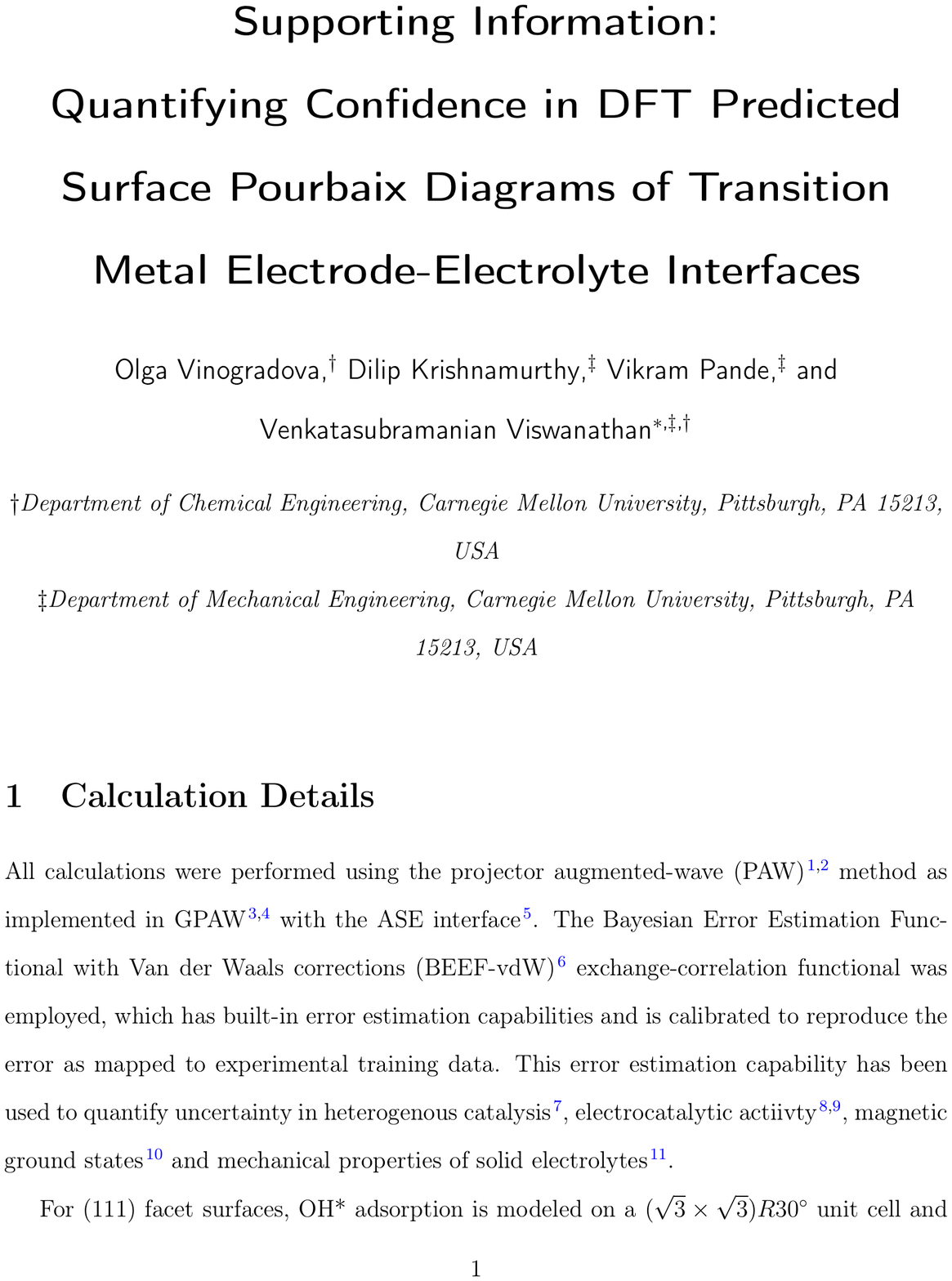}
\end{document}